\documentclass[11pt,a4paper,english,amssymb,nofootinbib,superscriptaddress]{revtex4}
\usepackage[T1]{fontenc}
\usepackage[latin9]{inputenc}
\usepackage{ifsym}
\usepackage{mathrsfs}
\usepackage{amsmath}
\usepackage{amssymb}
\usepackage{esint}
\usepackage[all]{xy}
\usepackage{graphicx}

\makeatletter
\@ifundefined{textcolor}{}
{%
 \definecolor{BLACK}{gray}{0}
 \definecolor{WHITE}{gray}{1}
 \definecolor{RED}{rgb}{1,0,0}
 \definecolor{GREEN}{rgb}{0,1,0}
 \definecolor{BLUE}{rgb}{0,0,1}
 \definecolor{CYAN}{cmyk}{1,0,0,0}
 \definecolor{MAGENTA}{cmyk}{0,1,0,0}
 \definecolor{YELLOW}{cmyk}{0,0,1,0}
}

\makeatother

\usepackage{babel}
\begin{document}

\title{Linearization Instability for Generic Gravity in AdS }

\author{Emel Altas}

\affiliation{Department of Physics,\\
 Middle East Technical University, 06800, Ankara, Turkey}
 
\author{Bayram Tekin}

\affiliation{Department of Physics,\\
 Middle East Technical University, 06800, Ankara, Turkey}
 
\date{\today}

\begin{abstract}

In general relativity, perturbation theory about a background solution
fails if the background spacetime has a Killing symmetry and a
compact spacelike Cauchy surface. This failure, dubbed as {\it linearization instability}, shows itself as non-integrability of the perturbative infinitesimal deformation to a finite deformation
of the background. Namely, the linearized field equations have spurious solutions which cannot be obtained from the linearization of exact solutions. In practice, one can show the failure of the linear perturbation theory  by showing that a certain quadratic (integral) constraint on the linearized solutions is not satisfied. For non-compact Cauchy surfaces, the situation is different and for example, Minkowski space, having a non-compact Cauchy surface, is linearization stable. Here we study the linearization instability in generic metric theories
of gravity where Einstein's theory is modified with additional curvature terms.
We show that, unlike the case of general relativity, for modified theories even in the non-compact
Cauchy surface cases, there are some theories which show linearization instability about their anti-de Sitter backgrounds. Recent $D$ dimensional critical and three dimensional chiral gravity theories are two such examples. This observation sheds light on the paradoxical behavior of vanishing conserved charges (mass, angular momenta) for non-vacuum solutions, such as black holes, in these theories. 

\end{abstract}

\maketitle

\section{INTRODUCTION}

There is a very interesting conundrum in nonlinear theories, such
as Einstein's gravity or its modifications with higher curvature terms:
exact solutions without symmetries (which are physically interesting) are hard to find, hence one resorts
to symmetric "background" solutions and develops a perturbative expansion about them. But it turns out
that exactly at the symmetric solutions, namely about solutions having Killing
vector fields, naive first order perturbation theory fails under certain conditions. The set of solutions to Einstein's equations forms a smooth manifold except at the solutions with infinitesimal symmetries and spacetimes with compact Cauchy surfaces for which there arise conical singularities in the solution space.  Namely, perturbation theory in non-linear theories can yield results which are simply wrong in the sense that {\it some} perturbative solutions cannot be obtained from the linearization of exact solutions. Roughly speaking, the process of first linearizing the field equations and then finding the solutions to those linearized equations; and the process of linearization of exact solutions to the non-linear equations can yield different results if certain necessary criteria are not met with regard to the background solution about which perturbation theory is carried out. Actually, the situation is more serious: linearized field equations can have spurious solutions which do not come from exact solutions. This could happen for various reasons and
the failure of the first order perturbation theory can be precisely
defined, as we shall do below.  Figure 1 summarizes the results. 

\begin{figure}
\begin{centering}
\includegraphics[scale=0.4]{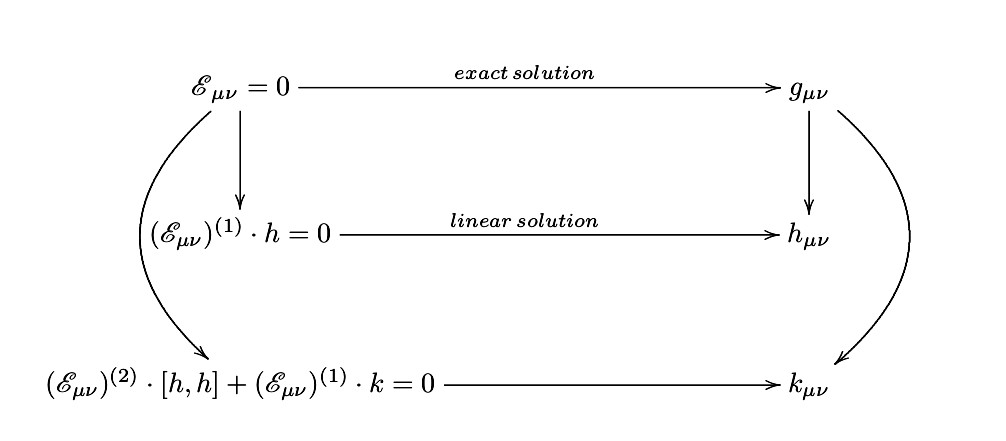} 
\par\end{centering}
\caption {The vertical straight arrows show first order linearization while the curved ones show second order linearization. For a linearization stable theory, the diagram makes sense and the solution to the linearized equation $h$ is not further restricted at the second order which means that there is a symmetric tensor $k$ that satisfies the second order equation in the bottom left. The details of the symbols are explained in the next section.}
\end{figure}

Let us give a couple of early observations
in this issue in the context of general relativity (GR) before we start the discussion in generic gravity.
One clear way to see the failure
of the perturbation theory is through the {\it initial value } formulation
of the theory for globally hyperbolic, oriented, time-orientable
spacetimes with  the topology ${\mathcal{M}} \approx \Sigma \times\mathbb{R}$, where $\Sigma$ is a spacelike Cauchy surface on which the induced Riemannian metric
$\gamma$ and the extrinsic curvature {\it K} (as well as matter content of the theory) are defined.~\footnote{It is also common to formulate the constraint equations in terms of $\gamma$ and a tensor density of weight 1 defined as  $\pi := \sqrt{\text{det} \gamma} ( K - \gamma\text{tr}_\gamma K )$ which is the conjugate momentum of the induced metric $\gamma$.} 
We shall consider the matter-free case through-out the paper. Since GR is nonlinear,
the initial data cannot be arbitrarily prescribed: they must satisfy
the so called Hamiltonian and momentum constraints $\Phi_{i}(\gamma,K)=0$ with $i \in \{1,2,3,4 \}$ in four dimensions. If a given initial data $(\bar{\gamma}, \bar{K})$  solving the constraints is not isolated, meaning the linearized constraint equations  $\delta \Phi_{i}(\bar \gamma,\bar K)\cdot [\delta \gamma, \delta K]=0$
allow {\it viable} linearized solutions $(\delta \gamma, \delta K)$, then the theory is said to be {\it linearization
stable} about the initial Cauchy data. Deser and
Brill \cite{Deser_Brill} showed that in GR with a compact Cauchy surface having the topology
of a 3-torus, there are strong constraints on the perturbations of the initial data. Any such perturbation leads to contradictions in the sense that bulk integrals of conserved mass and angular momenta do not vanish, while since there is no boundary, they must vanish in this compact space: hence the background is an isolated solution. Put in another way, the linearized field equations about the background have solutions which
do not come from the linearization of exact solutions. This happens because, as
we shall see below, the linearized equations of the theory are not  {\it sufficient }
to constrain the linearized solutions: quadratic constraints on the linearized solutions, in the form of an
integral (so called {\it Taub conserved quantity} first introduced in \cite{Taub} for each Killing vector field), arise.

Most of the work regarding the
linearization stability or instability in gravity has been in the
context of GR with or without matter and with compact or with non-compact
Cauchy surfaces. A nice detailed account of all these in the context of GR is given in the book \cite{book}. See also \cite{ch1} where a chapter is devoted to this issue and the Taub conserved quantity construction which is not widely known in the physics community. Our goal here is to extend the discussion to generic gravity
theories: we show that if the field equations of the theory are defined by the Einstein tensor
plus a covariantly conserved two tensor, then a new source of linearization
instability that does not exist in GR arises, especially in de Sitter
or Anti-de Sitter backgrounds, with non-compact  spatial-surfaces. This happens because in these backgrounds
there are special {\it critical} points in the space
of parameters of the theory which conspire to cancel the conserved
charge (mass, angular momentum {\it etc}) of non-perturbative objects (black
holes) or the energies of the perturbative excitations. One needs to understand the origin
of this rather interesting phenomenon that non-vacuum objects have
the same charges as the vacuum. To give an example of this phenomenon let
us note that this is exactly what happens in chiral gravity  \cite{Strom1,Strom2,Strom3,Carlip}
in 2+1 dimensions where the Einstein tensor is augmented with the Cotton
tensor and the cosmological constant times the metric (namely a special limit of the cosmological topologically massive gravity \cite{djt}). In AdS,
at the chiral point, the contribution of the Cotton tensor and the
Einstein tensor cancel each other at the level of the conserved
charges. Exactly at that point, new ghost-like solutions, the so called
log modes arise \cite{Grumiller} and if the boundary conditions are not those of Brown-Henneaux type {\cite{BH}, then these modes are present in the theory with negative energies. This would mean that the theory has no vacuum. But it was argued
in \cite{Strom2,Carlip} that chiral gravity in AdS has a linearization
instability which would remedy this problem. A similar phenomenon occurs in critical gravity in all
dimensions \cite{Pope, Tahsin}. Here we give a systematic discussion
of the linearization stability and instability in generic gravity
theories and study these two theories as examples. We will not  follow the route of defining the theory in the 3+1
setting and considering the instability problem on the Cauchy data. The reason for this is the following: in GR for asymptotically flat spacetimes, splitting the problem into the constraints on the Cauchy data and the evolution of the 3-metric and the extrinsic curvature turns the stability problem to a problem in elliptic  operator theory which is well-developed and sufficient to rigorously prove the desired results.  In the initial value formulation setting, the problem becomes a problem of determining the surjectivity of a linear operator, namely the linearized constraint operator. But this method is not convenient for our purposes, since the source of the linearization instability in the extended gravity models that we shall discuss is quite different and so the full spacetime formulation is  much better-suited for our problem. In GR, as noted in the abstract, what saves the Minkowski space from the linearization instability is its non-compact Cauchy surfaces  as was shown by Choquet-Bruhat and Deser \cite{Choquet-Bruhat}. This result is certainly consistent with the non-zero conserved charges (ADM mass or angular momentum) that can be assigned to an asymptotically-flat 3 dimensional Cauchy surface.

The layout of the paper is as follows: In section II, we discuss the linearization stability in generic gravity theory and derive the  second order constraints on the solutions of the linearized field equations. Of  course these constraints are all related to the diffeomorphism invariance and the Bianchi identities of the theory. Hence we give a careful discussion of the linearized forms of the field equations and their gauge invariance properties. As the second order perturbation theory about a generic background is quite cumbersome in the local coordinates, we carry out the index-free computations in the bulk of the paper and relegate some parts of the component-wise computations to the appendices. In section II, we establish the relation between the Taub conserved quantities coming from the second order  perturbation theory and the Abbott-Deser-Tekin (ADT) charges coming from the first order perturbation theory.  We study the linearization stability and instability of the Minkowski space, chiral gravity and critical gravity as examples. In a forth-coming paper, we shall give a more detailed analysis of the chiral gravity discussion in the initial value formulation context.

\section{Linearization Stability in Generic Gravity}

Let us consider the matter-free equation of a generic gravity theory in a $D$-dimensional spacetime, whose dynamical field is the metric tensor $g$ only. In the index-free notation the covariant two-tensor equation reads
\begin{equation}
\text{\ensuremath{\mathscr{E}}}(g)=0,
\label{gen_theory}
\end{equation}
together with the  covariant divergence condition which comes from the diffeomorphism invariance of the theory
\begin{equation}
\delta_{g}\text{\ensuremath{\mathscr{E}}}(g)=0,
\end{equation}
where $\delta_{g}$ denotes the divergence operator with respect to the metric $g$. (As usual, one uses the musical isomorphism to extend the divergence from the contravariant tensors to the covariant ones.) Here we generalize the discussion in \cite{Marsden_Fischer,Marsden} given for Einstein's theory to generic gravity. Let us assume that there
is a one-parameter family of solutions to (\ref{gen_theory}) denoted as $g(\lambda)$
which is at least twice differentiable with respect to $\lambda$ parameterizing the solution set. Then we can explore the consequences
of this assumption with the help of the following identifications : 
\begin{equation}
\bar{g}:=g(\lambda)\bigg\rvert_{\lambda=0}, \hskip 0.7 cm h:=\frac{d}{d\lambda}g(\lambda)\bigg\rvert_{\lambda=0},\hskip 0.7 cm k:=\frac{d^{2}}{d\lambda^{2}}g(\lambda)\bigg\rvert_{\lambda=0}.
\end{equation}
At this stage there is of course no immediate relation between the
two covariant tensor fields $h$ (the first derivative of the metric)  and $k$ (the second derivative of the metric) but, as we shall see later, consistency of the theory, {\it i.e.} the first order and the second order linearized forms of the field equations will relate them. We would first like to find that relation. 

We assume that $\bar{g}$
exactly solves the vacuum equations, $\text{\ensuremath{\mathscr{E}}}(\bar{g})=0$, and we compute the
first derivative of the field equations with respect to $\lambda$ and evaluate it at $\lambda =0$
as
\begin{equation}
\frac{d}{d\lambda}\text{\ensuremath{\mathscr{E}}}(g(\lambda))\bigg\rvert_{\lambda=0}=D\text{\ensuremath{\mathscr{E}}}(g(\lambda))\cdot\frac{dg(\lambda)}{d\lambda}\bigg\rvert_{\lambda=0}=0,
{\label{first_lin}}
\end{equation}
where $D$ denotes the Fr\'echet derivative and the center-dot denotes "along the direction of the tensor that comes next" and we have used the chain rule. In local coordinates, this equation is just the first order "linearization"
of the field equations (\ref{gen_theory}) which we shall denote as $\left(\text{\ensuremath{\mathscr{E}}}_{\mu\nu}\right)^{(1)}\cdot h=0$.
It is important to understand that solutions of (\ref{first_lin}) yield all possible $h$ tensors (up to diffeomorphisms),
which are tangent to the exact solution $g(\lambda)$ at $\lambda=0$ in the space of solutions. To understand if there are any {\it  further } constraints on the linearized solutions $h$, let us consider
the second derivative of the field equation with respect to $\lambda$
and evaluate it at $\lambda=0$ to arrive at 
\begin{equation}
\frac{d^{2}}{d\lambda^{2}}\text{\ensuremath{\mathscr{E}}}\bigg(g(\lambda)\bigg)\bigg\rvert_{\lambda=0}=\Bigg (D^{2}\text{\ensuremath{\mathscr{E}}}(g(\lambda))\cdot\left[\frac{dg(\lambda)}{d\lambda},\frac{dg(\lambda)}{d\lambda}\right]+D\text{\ensuremath{\mathscr{E}}}(g(\lambda))\cdot\frac{d^{2}g(\lambda)}{d\lambda^{2}}\Bigg)\bigg\rvert_{\lambda=0}=0,
 \label{main1}
\end{equation}
where we have used the common notation for the second Fr\'echet derivative in the first term and employed the chain rule when needed.  We can write (\ref{main1}) in local coordinates as
\begin{equation}
(\text{\ensuremath{\mathscr{E}}}_{\mu\nu})^{(2)}\cdot [h,h]+\left(\text{\ensuremath{\mathscr{E}}}_{\mu\nu}\right)^{(1)}\cdot k=0,
{\label{main2}}
\end{equation}
where  again $(\text{\ensuremath{\mathscr{E}}}_{\mu\nu})^{(2)}\cdot [h,h]$
denotes the second order linearization of the field equations about the background $\bar{g}$. Even though
this equation is rather simple, it is important to understand its
meaning to appreciate the rest of the discussion. This is the equation given in the bottom-left corner of Figure 1. Given a solution $h$ of   $\left(\text{\ensuremath{\mathscr{E}}}_{\mu\nu}\right)^{(1)}\cdot h=0,$ equation (\ref{main2})   {\it determines } 
the tensor field $k$, which is the second order derivative of the
metric $g(\lambda)$ at $\lambda=0.$ If such a $k$ can be found then
there is no further constraint on the linearized solution $h$. In that case, the field equations are said to be linearization stable at the exact solution $\bar{g}$. This says that the infinitesimal deformation $h$ is tangent to a full (exact) solution and hence it is integrable to a full solution.  Of course, what is tacitly assumed here is that, in solving for $k$ in (\ref{main2}), one cannot change the first order solution $h$, it must be kept intact for the perturbation theory to make any sense.  

We can understand these results from a more geometric vantage point as follows. For the spacetime manifold ${\cal{M}}$, let ${\mathcal{S}}$ denote the set of solutions of the field equations $\text{\ensuremath{\mathscr{E}}}(g)=0$. The obvious question is (in a suitable Sobolev topology), when does this set of solutions form a smooth manifold whose tangent space at some "point" $\bar{g}$ is the space of solutions ($h$) to the linearized equations?  The folklore in the physics literature is not to worry about this question and just assume that the perturbation theory makes sense and the linearized solution can be improved to get better solutions, or the linearized solution is assumed to be integrable to a full solution. But as we have given examples above, there are cases when the perturbation theory fails and the set ${\mathcal{S}}$ has a conical singularity instead of being a smooth manifold. 
One should not confuse this situation with the case of dynamical instability as the latter really allows a "motion" or perturbation about a given solution. Here linearization instability refers to a literal break-down of the first order perturbation theory.   
It is somewhat a non trivial matter to show that there are no further constraints  beyond the second order perturbation theory: In Einstein's gravity, this is related to the fact that constraint equations are related to zeros of the moment maps \cite{Marsden_lectures}. For generic gravity, this issue deserves to be further studied. 

\subsection{Taub conserved quantities and ADT charges}

So far, in our discussion we have not assumed anything about whether the spacetime has
a compact Cauchy surface or not. First, let us now assume that the spacetime
has a compact spacelike Cauchy surface and at least one Killing vector
field. Then we can get an \emph{integral constraint} on $h$, without
referring to the $k$ tensor as follows. Let $\bar{\text{\ensuremath{\xi}}}$
be a Killing vector field of the metric $\bar{g}$, then the following vector field \footnote{ For the lack of a better notation, note that $\bar{\text{\ensuremath{\xi}}}$ is contracted with the covariant background tensor with a center dot which we shall employ in what follows and it should not be confused with the center dot in the Fr\'echet derivative.}
\begin{equation}
T:=\bar{\text{\ensuremath{\xi}}}\cdot D^{2}\text{\ensuremath{\mathscr{E}}}(\bar{g})\cdot\left[h,h\right],
\end{equation}
is divergence free, since $\delta_{\bar{g}}D^{2}\text{\ensuremath{\mathscr{E}}}(\bar{g}).\left[h,h\right]=0$
due to the linearized Bianchi identity . Then we can integrate $T$
over a compact hypersurface $\Sigma$ and observe that the integral (for the sake of definiteness, here we consider the 3+1 dimensional case) 
\begin{equation}
\intop_{\Sigma}d^{3}\Sigma\thinspace\sqrt{\gamma}\thinspace{T}\cdot\hat{n}_{\Sigma}
\end{equation}
is independent of the hypersurface $\Sigma$ where $\gamma$
is the pull-back metric on the hypersurface and $\hat{n}_{\Sigma}$ is the
unit future pointing normal vector. Let us restate the result in a
form that we shall use below: given {\it two} compact  disjoint hypersurfaces 
$\Sigma_{1}$
and $\Sigma_{2}$ in the spacetime ${\mathcal{M}}$, we have the statement of the  "charge conservation"  as the equality of the integration over the two hypersurfaces 
\begin{equation}
\intop_{\Sigma_{1}}d^{3}\thinspace\Sigma_{1}\thinspace\sqrt{\gamma_{\Sigma_{1}}}\thinspace{T}\cdot\hat{n}_{\Sigma_{1}}=\intop_{\Sigma_{2}}d^{3}\thinspace\Sigma_{2}\thinspace\sqrt{\gamma_{\Sigma_{2}}}\thinspace{T}\cdot\hat{n}_{\Sigma_{2}}.
\end{equation}
 We can now go to (\ref{main2}) and after contracting it with the Killing
tensor $\bar{\text{\ensuremath{\xi}}}$, and integrating over  $\Sigma$, we obtain the identity 
\begin{equation}
\intop_{\Sigma} d^{3}\Sigma\thinspace\sqrt{\gamma}\thinspace\bar{\text{\ensuremath{\xi}}}^{\mu}\hat{n}^{\nu}(\text{\ensuremath{\mathscr{E}}}_{\mu\nu})^{(2)}\cdot [h,h]\thinspace=-\intop_{\Sigma} d^{3}\Sigma\thinspace\sqrt{\gamma}\thinspace\bar{\text{\ensuremath{\xi}}}^{\mu}\hat{n}^{\nu}\left(\text{\ensuremath{\mathscr{E}}}_{\mu\nu}\right)^{(1)}\cdot k\thinspace.
\label{denklik}
\end{equation}
Let us study the right-hand side more carefully. In a generic theory,
this conserved Killing charge is called the Abbott-Deser-Tekin (ADT)
charge when the symmetric two-tensor $k$ is just the linearized
two tensor $h$ \cite {Abbott_Deser, Deser_Tekin}. Once the field equations of the theory are given, it
is possible, albeit after some lengthy computation, to show that one
can write the integral on the right-hand side as a total derivative:
\begin{equation}
\bar{\text{\ensuremath{\xi}}}^{\mu}\left(\text{\ensuremath{\mathscr{E}}}_{\mu\nu}\right)^{(1)}\cdot h=\bar{\nabla}_{\alpha}\left(\text{\ensuremath{\mathscr{F}}}^{\alpha}\,_{\nu\mu}\bar{\text{\ensuremath{\xi}}}^{\mu}\right),
\label{div_denk}
\end{equation}
with an anti-symmetric tensor $\mathscr{F}$ in $\alpha$ and $\nu$.  
Hence if the Cauchy surface is compact without a boundary, the ADT charge
vanishes identically, namely
\begin{equation}
Q_{ADT}\left[\bar{\text{\ensuremath{\xi}}}\right] :=\intop_{\Sigma} d^{3}\Sigma\thinspace\sqrt{\gamma}\thinspace\hat{n}^{\nu}\thinspace\bar{\text{\ensuremath{\xi}}}^{\mu}\left(\text{\ensuremath{\mathscr{E}}}_{\mu\nu}\right)^{(1)}\cdot h=0,
\label{ADT}
\end{equation}
which via (\ref{denklik}) says that one has the vanishing of the integral on
the left hand-side which is called the Taub conserved quantity:
\begin{equation}
Q_{Taub}\left[\bar{\text{\ensuremath{\xi}}}\right]:=\intop_{\Sigma} d^{3}\Sigma\thinspace\sqrt{\gamma}\thinspace\hat{n}^{\nu}\thinspace\bar{\text{\ensuremath{\xi}}}^{\mu}\thinspace(\text{\ensuremath{\mathscr{E}}}_{\mu\nu})^{(2)}\cdot [h,h]=0,
\label{ttt}
\end{equation}
which must be {\it automatically} satisfied for the case when $h$ is an
integrable deformation. Otherwise this equation is a second order constraint on
the linearized solutions. Even though the ADT potential $\mathscr{F}$
was explicitly found for a large family of gravity theories, such as Einstein's
gravity \cite{Abbott_Deser}, quadratic gravity \cite{Deser_Tekin}, $f(Riem)$ theories \cite{SST}, and some examples will be given below, we can still refine
the above argument of the vanishing of both the ADT and Taub conserved quantities
without referring to the ADT potential (or more explicitly without
referring to (\ref{div_denk})). The following argument was given for Einstein's
gravity in \cite{Marsden} which immediately generalizes to the most general
gravity as follows: consider the ADT charge (\ref{ADT}) and assume that in the spacetime one
has two disjoint compact hypersurfaces $\Sigma_{1}$ and $\Sigma_{2}$ as above.
Then the statement of conservation of the charge is simply
\begin{equation}
Q_{ADT}\left(\bar{\text{\ensuremath{\xi}}},\Sigma_{1}\right)=Q_{ADT}\left(\bar{\text{\ensuremath{\xi}}},\Sigma_{2}\right).
\end{equation}
Now let $k$ be a two tensor which is $k_{1}$ and non-zero on $\Sigma_{1}$
and $k_{2}$ and zero near $\Sigma_{2}$, then $Q_{ADT}\left(\bar{\text{\ensuremath{\xi}}},\Sigma_{2}\right)=0$
so $Q_{ADT}\left(\bar{\text{\ensuremath{\xi}}},\Sigma_{1}\right)=0$
which in turn yields the vanishing of the Taub conserved quantities via (\ref{denklik}).

To summarize the results obtained so far, let us note that assuming
an integrable infinitesimal deformation $h$, which is by definition a solution
to the linearized field equations about a background $\bar{g}$ solution,
we arrived at  (\ref{main2}). And the discussion after that equation
showed that Taub conserved quantities constructed with a Killing vector field, from the second order linearization,
$(\text{\ensuremath{\mathscr{E}}}_{\mu\nu})^{(2)}\cdot [h,h]$,
and the ADT charges constructed from the first order linearization,
$(\text{\ensuremath{\mathscr{E}}}_{\mu\nu})^{(1)}\cdot h$,
vanish identically for the case of compact Cauchy hypersurfaces without
a boundary. If these integrals do not vanish, then there is a contradiction  and the linearized solution 
$h$  is further constraint. Hence it is not an integrable deformation, namely, $h$ is not in the tangent space about the point $\bar{g}$ in the space of solutions.  For Einstein's theory with compact Cauchy surfaces, it was shown that the necessary condition for linearization  stability is the absence of Killing vector fields \cite{Moncrief,Marsden_Arms}. As noted above, the interesting issue is that further study reveals that besides the quadratic constraint, there are no other constraints on the solutions to the linearized equations \cite{Marsden_lectures}.

\subsection{Gauge Invariance of the charges}

Of course there is one major issue that we still must
address that is the gauge-invariance (or coordinate independence)
of the above construction which we show now. Following \cite{Marsden}, first let us consider a (not necessarily small) diffeomorphism
$\varphi$ of the spacetime as $\varphi:{\mathcal{M}}\rightarrow {\mathcal{M}}$. Then we
demand that having obtained our rank two tensor $\text{\ensuremath{\mathscr{E}}}(g)$
from a diffeomorphism invariant action (or from a diffeomorphism invariant
action up to a boundary term as in the case of topologically massive gravity) we have a global statement of diffeomorphism invariance as 
\begin{equation}
\text{\ensuremath{\mathscr{E}}}\left(\varphi^{*}g\right)=\varphi^{*}\text{\ensuremath{\mathscr{E}}}\left(g\right),
\label{global}
\end{equation}
 which states that $\text{\ensuremath{\mathscr{E}}}$ evaluated
for the pull-back metric is equivalent to the pull-back of $\text{\ensuremath{\mathscr{E}}}$
evaluated for $g.$ Let us now consider a one-parameter family of
diffeomorphisms as $\varphi_\lambda$, generated by a vector field
${X}$ well-defined on some region of the spacetime. Let
$\varphi_{0}$ be the identity diffeomorphism denoted as $\varphi_{0}=I_{\mathcal{M}}.$
Then we can differentiate (\ref{global}) with respect to $\lambda$ once to get
\begin{equation}
\frac{d}{d\lambda}\text{\ensuremath{\mathscr{E}}}\left(\varphi_{\lambda}^{*}{\text{}}g\right)=\frac{d}{d\lambda}\varphi_{\lambda}^{*}\text{\ensuremath{\mathscr{E}}}\left(g\right),
\end{equation}
which, after making use of the chain rule, yields
\begin{equation}
D\text{\ensuremath{\mathscr{E}}}\left(\varphi_{\lambda}^{*}{\text{}}g\right)\cdot\frac{d}{d\lambda}\varphi_{\lambda}^{*}{\text{}}g=\varphi_{\lambda}^{*}{\text{}}\bigg (\text{\ensuremath{\mathscr{L}}}_{X}\text{\ensuremath{\mathscr{E}}}\left(g\right)\bigg),
\label{onyedi}
\end{equation}
where $\text{\ensuremath{\mathscr{L}}}_{X}$ is the Lie derivative
with respect to the vector field $X$. Taking the derivative of the last equation with respect to $g$ yields
\begin{equation}
D^{2}\text{\ensuremath{\mathscr{E}}}(g)\cdot\bigg(h,\text{\ensuremath{\mathscr{L}}}_{X}g\bigg)+D\text{\ensuremath{\mathscr{E}}}(g)\cdot\text{\ensuremath{\mathscr{L}}}_{X}h=\text{\ensuremath{\mathscr{L}}}_{X}\bigg(D\text{\ensuremath{\mathscr{E}}}\left(g\right)\cdot h\bigg).
\label{onsekiz}
\end{equation}
In components, and after setting $\lambda=0$, equation (\ref{onyedi}) reads, respectively
\begin{equation}
\delta_{X}\left(\text{\ensuremath{\mathscr{E}}}_{\mu\nu}\right)^{(1)}\cdot h=\text{\ensuremath{\mathscr{L}}}_{X}{\text{\ensuremath{\mathscr{E}}}}_{\mu\nu}(\bar{g}),
\label{birinci}
\end{equation}
and equation (\ref{onsekiz}) reads
\begin{equation}
\delta_{X}\left(\text{\ensuremath{\mathscr{E}}}_{\mu\nu}\right)^{(2)}\cdot \left[h,h\right]+\left(\text{\ensuremath{\mathscr{E}}}_{\mu\nu}\right)^{(1)}\cdot \text{\ensuremath{\mathscr{L}}}_{X}h=\text{\ensuremath{\mathscr{L}}}_{X}\left(\text{\ensuremath{\mathscr{E}}}_{\mu\nu}\right)^{(1)}\cdot h,
\label{ikinci}
\end{equation}
where $\delta_{X}\left(\text{\ensuremath{\mathscr{E}}}_{\mu\nu}\right)^{(1)}\cdot h$
denotes the variation of the background tensor $\left(\text{\ensuremath{\mathscr{E}}}_{\mu\nu}\right)^{(1)}\cdot h$
under the flow of $X$ or under the infinitesimal diffeomorphisms. Since
${\text{\ensuremath{\mathscr{E}}}}_{\mu\nu}(\bar{g})=0$, (\ref{birinci}) says that
$\left(\text{\ensuremath{\mathscr{E}}}_{\mu\nu}\right)^{(1)}\cdot h$
is gauge invariant: $\delta_{X}\left(\text{\ensuremath{\mathscr{E}}}_{\mu\nu}\right)^{(1)}\cdot h=0$.
Similarly (\ref{ikinci}) yields 
\begin{equation}
\delta_{X}\left(\text{\ensuremath{\mathscr{E}}}_{\mu\nu}\right)^{(2)}\cdot \left[h,h\right]+\left(\text{\ensuremath{\mathscr{E}}}_{\mu\nu}\right)^{(1)}\cdot \text{\ensuremath{\mathscr{L}}}_{X}h=0,
\label{ucuncu}
\end{equation}
since $\left(\text{\ensuremath{\mathscr{E}}}_{\mu\nu}\right)^{(1)}\cdot h=0$
by assumption, the right hand side of (\ref{ikinci}) vanishes. It is worth stressing that since generically  $\left(\text{\ensuremath{\mathscr{E}}}_{\mu\nu}\right)^{(1)}\cdot \text{\ensuremath{\mathscr{L}}}_{X}h$ is not zero,  the second order expansion $\left(\text{\ensuremath{\mathscr{E}}}_{\mu\nu}\right)^{(2)}\cdot \left[h,h\right]$ is not gauge invariant but transforms according to (\ref{ucuncu}).
Gauge invariance of the Taub conserved quantity and the ADT charge
follows immediately from (\ref{ucuncu}). Contracting that equation with the Killing vector
field $\bar{\text{\ensuremath{\xi}}}$ and integrating over the Cauchy
surface, one finds 
\begin{equation}
\intop_{\Sigma} d^{3}\Sigma\thinspace\sqrt{\gamma}\thinspace{n^{\nu}}\left[\bar{\text{\ensuremath{\xi}}}^{\mu}\delta_{X}\left(\text{\ensuremath{\mathscr{E}}}_{\mu\nu}\right)^{(2)}\cdot \left[h,h\right]+\bar{\text{\ensuremath{\xi}}}^{\mu}\left(\text{\ensuremath{\mathscr{E}}}_{\mu\nu}\right)^{(1)}\cdot \text{\ensuremath{\mathscr{L}}}_{X}h\right]=0.
\end{equation}
Since we have already shown that the second term can be written as a
divergence we can drop it out, the remaining part is the Taub conserved quantity which is shown to be is gauge invariant, by this construction. The
above discussion has been for a generic gravity theory based on the metric tensor as the only dynamical field, let us consider Einstein's gravity as an explicit example.

\subsection{Linearization Stability in Einstein's Gravity}

Let {\it Ein} denote the $(0,2)$ Einstein tensor, and $h$ denote a symmetric
two tensor field as described above and $X$ be a vector field, then the effect of infinitesimal one-parameter diffeomorphisms generated by $X$ follows as
\begin{equation}
DEin(g)\cdot\text{\ensuremath{\mathscr{L}}}_{X}g=\text{\ensuremath{\mathscr{L}}}_{X}Ein(g),
\end{equation}
which in local coordinates reads
\begin{equation}
\delta_{X}\left(G_{\mu\nu}\right)^{(1)}\cdot h=\text{\ensuremath{\mathscr{L}}}_{X}\bar{G}_{\mu\nu},
\end{equation}
where $G_{\mu\nu}:=Ein(e_{\mu},e_{\nu})$ and $Ein:=Ric-\frac{1}{2}Rg.$  We have already given the proof of the above equation for a generic theory in the previous part, but it pays to do it more explicitly in Einstein's theory: so it follows as
\begin{equation}
\delta_{X}(G_{\mu\nu})^{(1)}\cdot h=\delta_{X}(R_{\mu\nu})^{(1)}\cdot h-\frac{1}{2}\bar{g}_{\mu\nu}\delta_{X}(R)^{(1)}\cdot h-\frac{1}{2}\bar{R}\delta_{X}h_{\mu\nu},
\end{equation}
which just comes from the definition of the linearized Einstein tensor.  Then one can rewrite the above expression as desired: 
\begin{equation}
\delta_{X}(G_{\mu\nu})^{(1)}\cdot h=\text{\ensuremath{\mathscr{L}}}_{X}\left(\bar{R}_{\mu\nu}-\frac{1}{2}\bar{g}_{\mu\nu}\bar{R}\right)=\text{\ensuremath{\mathscr{L}}}_{X}\bar{G}_{\mu\nu}.
\end{equation}
At the second order of linearization, one has 
\begin{equation}
D^{2}Ein(g)\cdot\left(h,\text{\ensuremath{\mathscr{L}}}_{X}g\right)+DEin(g)\cdot\text{\ensuremath{\mathscr{L}}}_{X}h=\text{\ensuremath{\mathscr{L}}}_{X}\bigg(DEin(g)\cdot h\bigg),
\end{equation}
whose local version reads 
\begin{equation}
\delta_{X}(G_{\mu\nu})^{(2)}\cdot [h,h]+(G_{\mu\nu})^{(1)}\cdot\text{\ensuremath{\mathscr{L}}}_{X}h=\text{\ensuremath{\mathscr{L}}}_{X}(G_{\mu\nu})^{(1)}\cdot h.
\end{equation}
The explicit proof of this expression is rather long, hence we relegate it to Appendix A.

Now let us study the linearization stability of a particular solution to Einstein's gravity with a cosmological constant. Let $\bar{g}$ solve
the cosmological Einstein's field equations then the equation relevant to the  study of linearization stability of this solution is (\ref{main2}) which now reads
\begin{equation}
(\text{\ensuremath{\mathcal{G}}}_{\mu\nu})^{(2)}\cdot [h,h]+\left(\text{\ensuremath{\mathcal{G}}}_{\mu\nu}\right)^{(1)}\cdot k=0,
\label{eins}
\end{equation}
where $\left(\text{\ensuremath{\mathcal{G}}}_{\mu\nu}\right)^{(1)}\cdot k $ is a simple object but the second order object $(\text{\ensuremath{\mathcal{G}}}_{\mu\nu})^{(2)}\cdot [h,h]$ is quite cumbersome. It is very hard to use this equation to
show that for a generic background $\bar{g}_{\mu\nu}$, a $k_{\mu \nu}$ can
be found or cannot be found that satisfies (\ref{eins}). Therefore one actually
resorts to a weaker (sufficiency) condition that the Taub charges vanish which, as
we have seen, results from integrating this equation after contracting with
a Killing vector field $\bar{\text{\ensuremath{\xi}}}^{\mu}.$ To set the
stage for generic gravity theories about their AdS backgrounds, let
us study (\ref{eins}) in AdS and flat spaces. In that case one can
plug an explicit ansatz as follows: assume that such a $k$ exists in the form
\begin{equation}
k_{\mu\nu}=a\, h_{\mu\beta}h_{\nu}^{\beta}+b\, hh_{\mu\nu}+\bar{g}_{\mu\nu}(c\, h_{\alpha\beta}^{2}+d\, h^{2}),
\label{acik}
\end{equation}
where $k:=k_{\mu\nu}\bar{g}^{\mu\nu}$ and $a,b,c, d$ are
constants to  be determined and all the raising and lowering
is done with the background AdS metric $\bar{g}.$  Here we shall work in $D$ spacetime
dimensions. Inserting $k_{\mu\nu}$ as given in (\ref{acik}) in $\left(\text{\ensuremath{\mathcal{G}}}_{\mu\nu}\right)^{(1)}\cdot k$,
and choosing $a=1$ and $b=-\frac{1}{2}$, one arrives at 
\begin{equation}
\left(\text{\ensuremath{\mathcal{G}}}_{\mu\nu}\right)^{(2)}\cdot  [h,h] + \left(\text{\ensuremath{\mathcal{G}}}_{\mu\nu}\right)^{(1)}\cdot k  =: K_{\mu\nu},
\end{equation}
where $K_{\mu\nu}$ is a tensor which must vanish if the background
is linearization stable. Its explicit form is worked out in Appendix B. Let us consider the transverse traceless gauge, and make use of the field equations and the linearized field equations: Namely let us
use $\bar{R}_{\mu\nu}=\frac{2\varLambda}{D-2}\bar{g}_{\mu\nu}$
and $\left(\text{\ensuremath{\mathcal{G}}}_{\mu\nu}\right)^{(1)}\cdot h=0,$ which in this gauge reads 
$\bar{\text{\textifsymbol[ifgeo]{48}}}h_{\mu\nu}=\frac{4\Lambda}{(D-1)(D-2)}h_{\mu\nu}$ to arrive at 
\begin{equation}
K_{\mu\nu}=\bar{\nabla}_{\alpha}H^{\alpha}\thinspace_{\mu\nu}+\frac{\Lambda}{D-2}\left(c(D-2)+\frac{1}{2}\right)\bar{g}_{\mu\nu}h_{\alpha\beta}^{2}-\frac{1}{4}\bar{\nabla}_{\nu}h^{\alpha\beta}\bar{\nabla}_{\mu}h_{\alpha\beta}-\frac{\Lambda D}{(D-1)(D-2)}h_{\mu\beta}h_{\nu}^{\beta},
\label{div1}
\end{equation}
where the divergence piece is given as 
\begin{multline*}
H^{\alpha}\thinspace_{\mu\nu}:=\frac{1}{2}\left(h^{\alpha\beta}\bar{\nabla}_{\beta}h_{\nu\mu}+h_{\beta\nu}\bar{\nabla}_{\mu}h^{\alpha\beta}+h_{\beta\mu}\bar{\nabla}_{\nu}h^{\alpha\beta}-h_{\mu\beta}\bar{\nabla}^{\alpha}h_{\nu}^{\beta}-h_{\mu\beta}\bar{\nabla}^{\beta}h_{\nu}^{\alpha}\right)\\
-\frac{1}{4}\bar{g}_{\mu\nu}h_{\sigma\beta}\bar{\nabla}^{\beta}h^{\sigma\alpha}+\left(c(2-D)-\frac{1}{2}\right)\delta_{\nu}^{\alpha}h^{\sigma\beta}\bar{\nabla}_{\mu}h_{\sigma\beta}+\left(c(D-2)+\frac{5}{8}\right)\bar{g}_{\mu\nu}h_{\sigma\beta}\bar{\nabla}^{\alpha}h^{\sigma\beta}.
\end{multline*}
In the transverse-traceless gauge, the coefficient $d$ is not fixed
and can be set to zero. $K_{\mu\nu}$ has a single parameter $c$, that
one can choose to fix the stability of the flat spacetime (which was
proven by \cite{Choquet-Bruhat} using the linearization of the
constraints on a non-compact Cauchy surface in Minkowski space). Before
looking at the flat space case, let us note that one has $\bar{\nabla}_{\mu}K^{\mu\nu}=0$
as expected. Let us consider the flat space with $\Lambda=0$ and
use the Cartesian coordinates so that $\bar{\nabla}_{\alpha}\rightarrow\partial_{\alpha}.$
The corresponding linearized field equations become
\begin{equation}
\partial^{2}h_{\mu\nu}=0,
 \label{basit}
\end{equation}
together with the gauge choices $\partial_{\mu}h^{\mu\nu}=0=h.$ The
general solution of (\ref{basit}) can be exactly constructed as a superposition
of plane-wave solutions, hence it suffices to study the linearized
stability of flat space against the plane-wave modes which we take
to be the real part of 
\begin{equation}
h_{\mu\nu}=\text{\ensuremath{\varepsilon}}_{\mu\nu}e^{ik\cdot x},
\label{plane}
\end{equation}
together with $k^{\mu}\text{\ensuremath{\varepsilon}}_{\mu\nu}=0,$
$\text{\ensuremath{\varepsilon}}_{\mu}^{\mu}=0$ and $k^{2}=0,$\footnote {In a compact space without a boundary, $k=0$ mode should also be considered, in that case one has the solution $h_{\mu \nu} =\varepsilon_{\mu \nu} (c_1 t +c_2)$  which gives rise to linearization instability \cite{Higuchi} for the case of the torus.} which follow from the gauge condition and (\ref{basit}). Evaluating $K_{\mu\nu}$ for
this solution, one arrives at 
\begin{equation}
K_{\mu\nu}=k_{\nu}k_{\mu}\text{\ensuremath{\varepsilon}}_{\alpha\beta}\text{\ensuremath{\varepsilon}}^{\alpha\beta}e^{ik\cdot x}\left(2c(D-2)+\frac{5}{4}\right),
\end{equation}
which vanishes for the choice 
\begin{equation}
c=-\frac{5}{8(D-2)}\,.
\end{equation}
So  (\ref{eins}) is satisfied for 
\begin{equation}
k_{\mu\nu}=h_{\mu\beta}h_{\nu}^{\beta}-\frac{5}{8(D-2)}\bar{g}_{\mu\nu}h_{\alpha\beta}^{2}
\end{equation}
and therefore there is no further constraint on the linearized solutions (\ref{plane})
and the Minkowski space is linearization stable. Next we move on to the quadratic gravity theory.

\subsection{Linearization instability beyond Einstein's theory}

One of the reasons that led us to study the linearization instability
in generic gravity theories is an observation made in \cite{Deser_Tekin} where the conserved charges of generic gravity theories for asymptotically AdS backgrounds were constructed.\footnote{For an earlier zero energy result in the context of asymptotically flat backgrounds for purely quadratic gravity in four dimensions, see \cite{zero_en}.}The observation was  that in AdS backgrounds, the conserved energy and angular momenta vanish in generic 
gravity theories for all asymptotically AdS solutions at some particular
values of the parameters defining the theory (in fact a whole section in that paper was devoted for the zero energy issue). 
This apparent infinite degeneracy of the vacuum for AdS spaces, is in sharp contrast to the flat space case where the unique zero energy is attained only by the Minkowski space, namely the classical ground state. Let us expound upon this a little more: for ${\it all}$ purely metric based
theories, the energy (mass) of the space-time that asymptotically
approaches the flat space at spatial infinity is given by the ADM
formula 
\begin{equation}
M_{ADM}=\frac{1}{\kappa}\,\oint_{\partial \Sigma}\,dS_{i}\,(\partial_{j}h^{ij}-\partial^{i}h^{j}\,_{j})\,.\label{admass}
\end{equation}
It is well-known that $M_{ADM}\ge0$,
which is known as the positive energy theorem \cite{Schoen,Witten}.
An important part of this theorem is that the vacuum, namely the
flat space-time with $M_{ADM}=0$, is unique (up to diffeomorphisms
of course) \cite{Brill_Deser,Brill_Deser_Fad}.
It should be also noted that, the ADM mass is defined in flat Cartesian
coordinates but it was shown to be coordinate invariant. Here
one must be very careful, if proper decaying conditions are not realized
for $h_{ij}$, ${\it any}$ (positive, negative, finite or divergent)
value of mass can be assigned to the flat space. It is exactly these
properties of the ADM formula that made it a useful tool in geometry:
without even referring to Einstein's equations, one can take (\ref{admass})
to be a geometric invariant of an asymptotically flat manifold, modulo some decaying conditions on the first and the second fundamental forms of the spacelike surface. 
 
Once one deviates from  asymptotic flatness, then as we have noted, for higher derivative theories 
there are critical points which seem to make the vacuum infinitely degenerate, namely, the corresponding mass formula assigns any solution of the theory  the same zero charge.  Naively, one can try to understand the meaning of vanishing charges
for non-vacuum solutions (namely, non-maximally symmetric solutions)
as follows: 
\begin{itemize}

\begin{item}
There is a confinement of the relevant perturbations (in the weak coupling), just-like
in QCD in the strong coupling regime of color charge; and so a non-vacuum solution such as the proton has zero total color charge, same as the vacuum.  In the case of QCD, perturbation theory might yield spurious states
that cannot freely exist, such as quarks, as also noted in \cite{Strom2}. In gravity confinement would mean the confinement of mass-energy or some other properties under consideration such as chirality. But this would be highly unphysical because if there are no other conserved charges to suppress the creation of confined mass, then the vacuum state of gravity would be infinitely degenerate and creating confined mass would cost nothing.  
\end{item}

\begin{item}

The second possibility is that perturbation theory about a given
background solution, be it the maximally symmetric vacuum or not, may simply fail to exist
just because the background solution is an isolated solution in the
solution space. Namely, the solution space may fail to be a smooth
manifold. 
\end{item}

\end{itemize}

In fact, as discussed above, linearization
of non-linear equations such as Einstein's gravity and Yang-Mill's
theory showed that naive first order perturbation theory fails generically
when the background has  a Killing symmetry. To be more specific we 
consider two recent examples: the chiral gravity in 2+1 dimensions
which is a special case of topologically massive gravity with a cosmological
constant and the critical gravity which is a specific example of quadratic
gravity in AdS. These examples can be easily extended, as the phenomenon
we discuss is quite generic and takes place whenever Einstein's theory
with a cosmological constant is modified with some higher curvature terms.

To see how perturbation theory can fail let us go back to the necessary condition (\ref{main2}) and contract it with the Killing vector $\bar{\text{\ensuremath{\xi}}}^{\mu}$ to obtain
\begin{equation}
\bar{\text{\ensuremath{\xi}}}^{\mu}\left(\text{\ensuremath{\mathscr{E}}}_{\mu\nu}\right)^{(2)}\cdot \left[h,h\right]+\bar{\text{\ensuremath{\xi}}}^{\mu}\left(\text{\ensuremath{\mathscr{E}}}_{\mu\nu}\right)^{(1)}\cdot k=0.
\end{equation}
In some modified gravity theories one finds that the second term can be written as 
\begin{equation}
\bar{\text{\ensuremath{\xi}}}^{\mu}\left(\text{\ensuremath{\mathscr{E}}}_{\mu\nu}\right)^{(1)}\cdot k=c(\alpha_i, \bar{R}) \bar{\nabla}_{\alpha}{\mathcal{F}}_{1}^{\alpha}\thinspace_{\nu}+\bar{\nabla}_{\alpha}{\mathcal{F}}_{2}^{\alpha}\thinspace_{\nu},
\label{bound}
\end{equation}
where $c(\alpha_i, \bar{R}) $ is a constant  determined by the parameters $\alpha_i$ of the theory as well as the curvature invariants (symbolically denoted above as $\bar{R}$) of the background metric.  ${\mathcal{F}}_{i}^{\alpha\nu}$ are antisymmetric background tensors. It turns out that for asymptotically AdS spacetimes ${\mathcal{F}}_{2}^{\alpha\nu}$ vanishes identically at the boundary as it involves higher derivative terms of the perturbation, while ${\mathcal {F}}_{1}^{\alpha\nu}$ need not vanish if there are not so fast decaying fields such as, for example, the Kerr-AdS black holes. On the other hand for the particular choice of the parameters $c(\alpha_i, \bar{R}) =0$, one arrives at the constraint that again the Taub charges must  vanish identically
\begin{equation}
Q_{Taub}[\bar{\text{\ensuremath{\xi}}}]=\oint_{\Sigma} d^{D-1}\thinspace\Sigma\thinspace\sqrt{\gamma}\thinspace\bar{\text{\ensuremath{\xi}}}^{\mu}(\text{\ensuremath{\mathscr{E}}}_{\mu\nu})^{(2)}\cdot [h,h]=0.
\label{nine1}
\end{equation}
But this time we have the additional non-trivial equation
\begin{equation}
\oint_{\Sigma} d^{D-1}\thinspace\Sigma\thinspace\sqrt{\gamma}\thinspace\bar{\text{\ensuremath{\xi}}}^{\mu}\left(\text{\ensuremath{\mathscr{E}}}_{\mu\nu}\right)^{(1)}\cdot h\neq0.
\label{nine2}
\end{equation}
In general it is very hard to satisfy these two conditions simultaneously for all solutions. Therefore some solutions to the linearized equations $h$ turn out to be not integrable to a full solution, hence the linearization instability of the AdS background in these critical theories. Let us stress that we have not assumed that the Cauchy surfaces are compact: this type of linearization instability arises even in the non-compact  case.

\subsection{Linearization instability in quadratic gravity}

The message we would like to convey is a rather universal one in all
generic higher derivative gravity theories, but for the sake of being
concrete and yet sufficiently general, we shall consider the quadratic
gravity theory with the action (in $D$ dimensions) 
\begin{eqnarray}
I=\int d^{D}\, x\sqrt{-g}\Bigg (\frac{1}{\kappa}(R -2 \Lambda_0)+\alpha R^{2}+\beta R_{\mu\nu}^{2}+\gamma(R_{\mu\nu\rho\sigma}^{2}-4R_{\mu\nu}^{2}+R^{2})\Bigg ),\label{action}
\end{eqnarray}
where the last term is organized into the Gauss-Bonnet form, which
vanishes identically for $D=3$ and becomes a surface term for $D=4$.
But for $D\ge5$, it contributes to the field equations with at most
second order derivatives in the metric, just like the Einstein-Hilbert
part. Conserved gravitational charges of this theory  in its asymptotically AdS backgrounds were constructed
in \cite{Deser_Tekin} following the background space-time
techniques developed in \cite{Abbott_Deser} which is an extension of the ADM approach \cite{adm}. For any theory
with a Lagrangian density ${\cal {L}}=\frac{1}{\kappa}(R-2\Lambda_{0})+f(R_{\sigma\rho}^{\mu\nu})$,
for a generic differentiable function $f$ of the Riemann tensor and its contractions, the conserved charges follow
from those of (\ref{action}), as shown in \cite{SST} since any such theory can be written as a quadratic theory with effective coupling constants as far as its energy properties and particle content are concerned \cite{Tekin_particle}. In what follows, we quote some of the computations done in \cite{Deser_Tekin} here to make the ensuing discussion complete. The field equations  that follow from (\ref{action}) are 
\begin{eqnarray}
 & {\mathcal{E}}_{\mu \nu}[g]= & \frac{1}{\kappa}(R_{\mu\nu}-\frac{1}{2}g_{\mu\nu}R)+2\alpha R\,(R_{\mu\nu}-\frac{1}{4}g_{\mu\nu}\, R)+(2\alpha+\beta)(g_{\mu\nu}\Box-\nabla_{\mu}\nabla_{\nu})R\nonumber \\
 &  & +2\gamma\Big\{ RR_{\mu\nu}-2R_{\mu\sigma\nu\rho}R^{\sigma\rho}+R_{\mu\sigma\rho\tau}R_{\nu}^{\sigma\rho\tau}-2R_{\mu\sigma}R_{\nu}^{\sigma}-\frac{1}{4}g_{\mu\nu}(R_{\tau\lambda\rho\sigma}^{2}-4R_{\sigma\rho}^{2}+R^{2})\Big\}\nonumber \\
 &  & +\beta\Box(R_{\mu\nu}-\frac{1}{2}g_{\mu\nu}R)+2\beta(R_{\mu\sigma\nu\rho}-\frac{1}{4}g_{\mu\nu}R_{\sigma\rho})R^{\sigma\rho}=0.\label{eom}
\end{eqnarray}
As we shall study the stability/instability of the non-flat maximally symmetric solution (or solutions), let 
$\bar{g}$ represent such a solution with the curvature  tensors normalized as
\begin{equation}
\bar{R}_{\mu \rho \nu \sigma}  = \frac{ 2 \Lambda}{ (D-1)(D-2)} \big ( \bar{g}_{\mu \nu}  \bar{g}_{\rho \sigma} -   \bar{g}_{\mu \sigma}  \bar{g}_{\rho \nu} \big ), \hskip 0.5 cm \bar{R}_{\mu \nu} = \frac{2 \Lambda}{D-2} \bar{g}_{\mu \nu}, \hskip 0.5 cm \bar{R} = \frac{ 2 D \Lambda}{ D-2}.
\end{equation}
The field equations reduce to a single quadratic equation :
\begin{equation}
\frac{\Lambda-\Lambda_{0}}{2\kappa}+k \Lambda^{2}=0,\qquad k \equiv\left(D\alpha+\beta\right)\frac{\left(D-4\right)}{\left(D-2\right)^{2}}+\gamma\frac{\left(D-3\right)\left(D-4\right)}{\left(D-1\right)\left(D-2\right)}.\label{quadratic}\end{equation}
For generic values of the parameters of the theory, of course, there may not be a real solution and so the theory may not posses a maximally symmetric vacuum, but here we assume that there is a real solution to this algebraic equation (so $ 8 \Lambda_0 k \kappa +1 \ge 0$)  and study the linearization stability of this solution, which we call the (classical) vacuum or the background.  One can then linearize the field equations (\ref{eom}) about the vacuum and get at the linear order  
\begin{equation}
c_1\,({\cal {G}}_{\mu\nu})^{(1)}+\left(2\alpha+\beta\right)\left(\bar{g}_{\mu\nu}\bar{\square}-\bar{\nabla}_{\mu}\bar{\nabla}_{\nu}+\frac{2\Lambda}{D-2}\bar{g}_{\mu\nu}\right)(R)^{(1)}+\beta\left(\bar{\square}(\mathcal{G}_{\mu\nu})^{(1)}-\frac{2\Lambda}{D-1}\bar{g}_{\mu\nu}(R)^{(1)}\right)=0,\label{Linearized_eom}\end{equation}
where the constant in front of the first term is 
 \begin{equation}
c_1\equiv\frac{1}{\kappa}+\frac{4\Lambda D}{D-2}\alpha+\frac{4\Lambda}{D-1}\beta+\frac{4\Lambda\left(D-3\right)\left(D-4\right)}{\left(D-1\right)\left(D-2\right)}\gamma,\label{eqc}\end{equation}
and the linearized (background) tensors read
\begin{equation}
({\cal {G}}_{\mu\nu})^{(1)}=(R_{\mu\nu})^{(1)}-\frac{1}{2}\bar{g}_{\mu\nu}(R)^{(1)}-\frac{2\Lambda}{D-2}h_{\mu\nu},
\end{equation}
 which is just the linearized cosmological Einstein's tensor given in terms of the linearized Ricci tensor and the linearized scalar curvature which can be computed to be
\begin{equation}\nonumber
(R_{\mu\nu})^{(1)}=\frac{1}{2}\Big (\bar{\nabla}^{\sigma}\bar{\nabla}_{\mu}h_{\nu\sigma}+\bar{\nabla}^{\sigma}\bar{\nabla}_{\nu}h_{\mu\sigma}-\bar{\square}h_{\mu\nu}-\bar{\nabla}_{\mu}\bar{\nabla}_{\nu}h\Big),\qquad (R)^{(1)}=-\bar{\square}h+\bar{\nabla}^{\sigma}\bar{\nabla}^{\mu}h_{\sigma\mu}-\frac{2\Lambda}{D-2}h.
\end{equation}
Given a background Killing vector $\bar{\xi}$, (there are $D(D+1)/2$ number of Killing vectors for maximally symmetric spaces and the arguments work for any one of these Killing vectors) if we had not truncated the expansion of the field equations at ${\mathcal{O}(h)}$ but collected all the non-linear terms on the right-hand side, we would have gotten 
\begin{equation}
\bar{\text{\ensuremath{\xi}}}^{\mu}\left(\text{\ensuremath{\mathscr{E}}}_{\mu\nu}\right)^{(1)}\cdot h := \bar{\xi}^\mu T_{\mu \nu}[h^2, h^3, ... h^n...],
\label{lin}
\end{equation}
where  $T_{\mu \nu}[h^2, h^3, ...h^n...] $ represents all the higher order terms (and if there is a matter source with compact support of the energy-momentum tensor, it also includes that).  The next step is the crucial step: as was shown in \cite{Deser_Tekin}, one can write (\ref{lin}) as a divergence of two pieces as described by (\ref{bound})
\begin{equation} 
\bar{\text{\ensuremath{\xi}}}^{\mu}\left(\text{\ensuremath{\mathscr{E}}}_{\mu\nu}\right)^{(1)}\cdot h=c \, \bar{\nabla}_{\alpha}{\mathcal{F}}_{1}^{\alpha}\thinspace_{\nu}+\bar{\nabla}_{\alpha}{\mathcal{F}}_{2}^{\alpha}\thinspace_{\nu},
\label{bound2}
\end{equation}
where the constant   $c_1$  given in (\ref{eqc}) is shifted due to the $\beta$ term as 
\begin{equation}
c \equiv  c_1 + \frac{4\Lambda}{(D-1)(D-2)}\beta.
\end{equation}
The explicit forms of the ${\mathcal F}_i^{\mu \rho}$ tensors are found to be 
\begin{eqnarray}
{\mathcal F}_1^{\mu \rho}= 2 \bar{\xi}_\nu \bar{\nabla}^{ [ \mu}h^{\rho ]\, \nu} 
+2 \bar{\xi}^{ [ \mu } \bar \nabla^{ \rho ] } h +2  h^{\nu [ \mu}\bar{\nabla}^{\rho ]} \bar{\xi}_\nu +2 \bar{\xi}^{[ \rho} \bar{\nabla}_{\nu}h^{\mu  ] \, \nu} +
h\bar{\nabla}^\mu \bar{\xi}^\rho,
\label{f1}
\end{eqnarray}
and
\begin{eqnarray}
{\mathcal{F}}_{2}^{\mu \rho} =&(2\alpha+\beta)\Bigg(2 \bar{\xi}^{[\mu}\bar{\nabla}^{\rho]}(R)^{(1)}+(R)^{(1)}\bar{\nabla}^{\mu}\,\bar{\xi}^{\rho}\Bigg)
+2\beta\Bigg(\bar{\xi}^{\sigma}\bar{\nabla}^{[\rho}({\cal {G}}^{\mu]}\,_{\sigma})^{(1)}+({\cal {G}}^{[\rho\sigma})^{(1)}\bar{\nabla}^{\mu ]}\bar{\xi}_{\sigma}\Bigg).
\end{eqnarray}
For asymptotically AdS spacetimes, ${\mathcal{F}}_{2}^{\mu \rho} $ vanishes at spatial infinity  due to the vanishing of both of $ (R)^{(1)}$ and $({\cal {G}}_{\mu\sigma})^{(1)}$. As discussed in the previous section, vanishing of the constant $c$ leads to two strong constraints (\ref{nine1}) and (\ref{nine2}) on the linearized solution $h$ which is a statement of the instability of the background solution.  
Note that, for this higher order theory, we have not assumed that the spatial hypersurface is compact. ( In fact, to be more accurate, AdS is not globally hyperbolic and does not  have a Cauchy surface but one can work in the double cover which does). 

The point at which $c=0$ is the point when the mass of the spin-2 massive mode also vanishes and further, assuming $4\alpha(D-1) + D\beta=0$, one can also decouple the massive spin-0 mode in this theory and arrive at the so called {\it critical gravity } defined in $D=4$  \cite{Pope} for generic $D$ in \cite{Tahsin}.
All these conditions are compatible with the existence of a maximally symmetric vacuum.  
For critical gravity, the apparent  mass and angular momenta of all black holes and perturbative excitations with asymptotically AdS conditions vanish.\footnote{The energy of the perturbative bulk excitations can be constructed using the Ostrogradsky Hamiltonian \cite{Tahsin}.} But as we have seen here, perturbation theory used for both the excitations and the construction of the conserved quantities does not work exactly at the critical point: namely, the theory for the AdS background is not linearization stable. At the chiral point, there arise exact log-modes in chiral gravity \cite{Alisah, GGST} which are of the wave type but they do not correspond to the linearized log-modes of \cite{Grumiller}.
  
Just for the sake of completeness, let us note that if $c \ne 0$, then  the perturbation theory makes sense and the conserved charges of the theory for any asymptotically AdS solutions (such as the Kerr-AdS black holes) are simply given in terms of the conserved charges of the same solution in Einstein's gravity as 
\begin{align}
\frac{Q_{\text{quad}}(\bar{\xi})}{Q_{\text{Einstein}}(\bar{\xi})} = -\beta m_g^2 ,\label{Quad_charge}
\end{align}
where $m_g$ is the mass of the spin-2 graviton given as 
\begin{equation}
 -\beta m_g^2=\frac{1}{\kappa}+\frac{4\Lambda(D\alpha+\beta)}{D-2}+\frac{4\Lambda\left(D-3\right)\left(D-4\right)}{\left(D-1\right)\left(D-2\right)}\gamma.
\end{equation}
In (\ref{Quad_charge}), $Q_{\text{Einstein}}(\bar{\xi})$ refers to (with $\kappa_{\text{Newton}}=1$)
the conserved charge (mass, angular momenta) in the cosmological
Einstein's theory.

\subsection{Linearization instability in chiral gravity}

A model of quantum gravity even in the simpler $2+1$ dimensional setting has been rather elusive. One of the 
latest promising proposals was the so called chiral gravity \cite{Strom1} which is a specific limit of topologically massive gravity (TMG) \cite{djt} with the asymptotically AdS boundary conditions. TMG, as opposed to Einstein's gravity has non-trivial local dynamics hence in this respect, it might be more relevant  to the four dimensional gravity both at the classical and quantum level. The crux of the arguments of the quantum version chiral gravity is that the bulk theory is dual to a unitary and chiral conformal field theory (CFT) on the two dimensional boundary, whose symmetry is known to be one of the two copies of the Virasoro algebra {\cite{BH}.  Finding the correct conformal field theory would amount to defining the quantum gravity via the AdS/CFT duality \cite{Maldacena}. But immediately after the proposal of chiral gravity, it was realized that the theory has arbitrarily negative energy log modes that appear exactly at the chiral point and not only the dual CFT is not unitary (but a logarithmic one), but apparently  chiral gravity does not have even a classical vacuum \cite{Grumiller}.  If true, this of course would be disastrous  for chiral gravity. But later it was argued in \cite{Strom2,Carlip} that chiral gravity has  linearization instability against these log modes in AdS: namely, these perturbative negative energy solutions do not actually come from the linearization of any exact solution. If that is the case, then linearization instability saves chiral gravity certainly at the classical level and perhaps at the quantum level. Here we give further arguments of the existence of linearization  
instability in chiral gravity. 

The field equations of topologically massive gravity \cite{djt} with a negative cosmological constant ($\Lambda := -\frac{1}{\ell^2}$) is 
\begin{equation}
R_{\mu\nu}-\frac{1}{2}g_{\mu\nu}R -\frac{1}{\ell^2} g_{\mu\nu}
+\frac{1}{\mu}C_{\mu\nu}=0,
\label{fieldeqns1}
\end{equation}
where  the Cotton tensor in terms of the anti-symmetric $\eta-$tensor and the covariant derivative of the Schouten tensor reads
\begin{equation}
C_{\mu\nu}=\eta_{\mu}\,^{\alpha\beta}\nabla_{\alpha}S_{\beta\nu}, \hskip 0.5 cm S_{\mu\nu}=R_{\mu\nu}-\frac{1}{4}g_{\mu\nu}R.
\end{equation}
The boundary theory has two copies of the Virasoro algebra \cite{BH} for asymptotically AdS  boundary conditions given as 
\begin{equation}
c_{R/L}=\frac{3\text{\ensuremath{l}}}{2G_{3}}\left(1 \pm\frac{1}{\mu\text{\ensuremath{l}}}\right), 
\label{central}
\end{equation}
and the bulk theory has a single helicity 2 mode with a mass-square 
\begin{equation}
 m_{g}^{2}=\mu^{2}-\frac{1}{l^{2}}.
\end{equation}
It was shown in \cite{Tek_Deser} that the contraction of the Killing vector $(\bar{\xi})$ with the linearized equations coming from (\ref{fieldeqns1}) yields 
\begin{equation}
\bar{\xi}^\mu\Bigg (({\mathcal{G}}_{\mu\nu})^{(1)}+\frac{1}{\mu}(C_{\mu\nu})^{(1)} \Bigg) 
= \, \bar{\nabla}_{\alpha}{\mathcal{F}}_{1}^{\alpha}\thinspace_{\nu} [\bar{\Xi}]+\bar{\nabla}_{\alpha}{\mathcal{F}}_{3}^{\alpha}\thinspace_{\nu}[\bar \xi],
\label{lineareq}
\end{equation}
where ${\mathcal{F}}^{\mu \rho}_1$ was given in (\ref{f1}) whereas one finds ${\mathcal{F}}^{\mu \rho}_3$ to be
\begin{eqnarray}
{\mathcal{F}}^{\mu \rho}_3 [\bar{\xi}] = 
\eta^{\mu  \rho \beta} \, ({\mathcal G}_{\nu\beta})^{(1)} \, \bar{\xi}^{\nu}
+ \eta^{\nu \rho \beta} \, ({\mathcal G}^{\mu}\,_{\beta})^{(1)} \, \bar{\xi}_{\nu}
+ \eta^{\mu\nu\beta} \, ({\mathcal G}^{\rho}\,_{\beta})^{(1)} \, \bar{\xi}_{\nu},
\label{f3}
\end{eqnarray}
where  a new (twisted) Killing vector ($\bar{\Xi}$) appears:
\begin{equation}
 \bar{\Xi}^{\alpha} := \bar \xi^\alpha+ \frac{1}{2 \mu}\eta^{ \alpha \beta\nu} \, 
\bar{\nabla}_{\beta} \, \bar{\xi}_{\nu}.
\end{equation}
The conserved charges of TMG for asymptotically AdS backgrounds read as an integral over the circle at infinity as
\begin{equation}
Q [\bar{\xi}] = \frac{1}{8 \pi G_3} \, \oint_{\partial {\cal M}} \,
dS_{i} \, \left( {\mathcal{F}}^{0 i}_1 [\bar{\Xi}] + \frac{1}{2 \mu} \,{\mathcal{F}}^{0 i}_3 [\bar{\xi}]
\right) .
\end{equation}

Once again for the asymptotically  AdS cases ${\mathcal{F}}_{3}^{\alpha}\thinspace_{\nu}[\bar \xi]$ vanishes identically on the boundary as it involves the linearized Einstein tensor at infinity. For generic values of $\mu$ and $\ell$, the first term, that is $ {\mathcal{F}}_{1}^{\alpha}\thinspace_{\nu} [\bar{\Xi}]$ gives the conserved charges for the corresponding Killing vector. But, for $\mu^2 \ell^2 =1$, as was shown in \cite{Serkay} the angular momentum and the energy of the rotating black hole solutions with the rotation parameter ($j$) and the mass ($m$) related as ($j = m \ell$) (the  extremal BTZ black hole) vanishes identically. This particular point was further studied in \cite{Strom1} where it was argued and conjectured that the theory, so called {\it chiral} gravity, as one of the central charges noted above (\ref{central}) becomes zero, makes sense both classically and quantum mechanically. 

Classically the theory should have a stable vacuum and  quantum mechanically, it should have a  dual healthy boundary conformal field theory. In \cite{Strom1} it was shown that all the bulk excitations have vanishing energy exactly at the chiral point. Later new log modes that were not accounted for were found in \cite{Grumiller} which violated the existence of a ground state (namely, these modes have arbitrarily large negative energy compared to the zero energy of the vacuum).  For further work on chiral gravity, see \cite{Wise,Porrati}.
In \cite{Strom2} and \cite{Carlip} it was argued that the AdS has linearization instability in chiral gravity against these log modes. Here, our construction lends support to these arguments.

For the sake of concreteness, let us consider the background metric as
\begin{equation}
\bar{g} = - \bigg(1+\frac{r^2}{\ell^2}\bigg) dt^2 + \frac{ dr^2}{1+\frac{r^2}{\ell^2}}+ r^2 d\phi^2,
\end{equation} 
then for $\bar{\xi} = ( -1, 0, 0)$, referring to the time-like energy Killing vector, one finds the twisted Killing vector to be
\begin{equation}
\Xi =  ( -1, 0,  - \frac{1}{\ell^2 \mu}).
\end{equation} 
For this $\Xi$ to be a time-like Killing vector for {\it all} $r$, including the boundary at $r \rightarrow \infty$, one can see that (excluding the trivial $\mu \rightarrow \infty$ case) one must set $\mu^2 \ell^2 =1$, which is the chiral gravity limit. To further see this chiral gravity limit, let us recast  ${\mathcal{F}}^{\mu \rho}_1 [\Xi]$ using the  the superpotential ${\mathcal{K}}^{\mu\alpha\nu\beta}$ is defined by \cite{Abbott_Deser}
\begin{eqnarray}
{\mathcal{K}}^{\mu\nu\alpha\beta} :=\frac{1}{2}\bigg(\bar{g}^{\mu\beta}\tilde h^{\nu\alpha}+\bar{g}^{\nu\alpha}\tilde h^{\mu\beta}-\bar{g}^{\mu\nu}\tilde h^{\alpha\beta}-\bar{g}^{\alpha\beta}\tilde h^{\mu\nu}\bigg),\hskip1cm\tilde h^{\mu\nu} :=h^{\mu\nu}-\frac{1}{2}\bar{g}^{\mu\nu}h,
\end{eqnarray}
which yields 
\begin{equation}
{\mathcal{F}}^{\mu \rho}_1 [\Xi] =  \Xi_\nu \bar{\nabla}_{\beta} {\mathcal{K}}^{\mu \rho \nu \beta}-
{\mathcal{K}}^{ \mu \sigma \nu \rho }\bar{\nabla}_{\sigma}\bar{\Xi}_\nu.
\end{equation}
For all asymptotically AdS solutions with the Brown-Henneaux boundary conditions, one can show that  
\begin{equation}
{\mathcal{F}}^{\mu \rho}_1 [\Xi] = \bigg(1- \frac{1}{\ell^2 \mu^2}\bigg){\mathcal{F}}^{\mu \rho}_1 [\bar{\xi}],
\end{equation}
which vanishes at the chiral point. So exactly at this point, there exist second order integral constraints on the linearized solutions as discussed in the previous section. The log-modes of \cite{Grumiller} do not satisfy these integral constraints and so fail to be integrable to full solutions.\footnote{See \cite{log}  for a nice compilation of possible applications  of logarithmic field theories in the context of holography and gravity. }   

Let us compute the value of the Taub conserved quantity for the log solution which was given in the background with the global coordinates for which the  metric reads
\begin{equation}
 ds^2 = \ell^2\big(-\cosh^2{\rho}\, d\tau^2 +\sinh^2{\rho}\,d\phi^2+d\rho^2\big).
\end{equation}
For the coordinates $u=\tau+\phi$, $v=\tau-\phi$, at exactly in the chiral point, one has the following additional solution 
\begin{equation}
\begin{aligned}
 h_{ \mu\nu} = \frac{\sinh{\rho}}{\cosh^3{\rho}}\,\big(\cos{(2u)}\,\tau-\sin{(2u)}\,\ln{\cosh{\rho}}\big)&\left( \begin{array}{ccc}
0 & 0 & 1 \\
0 & 0 & 1 \\
1 & 1 & 0
\end{array} \right)_{\mu\nu}  \\
-\tanh^2\!\!{\rho}\,\big(\sin{(2u)}\,\tau+\cos{(2u)}\,\ln{\cosh{\rho}}\big)&\left( \begin{array}{ccl}
1 & 1 & \qquad \qquad 0 \\
1 & 1 & \qquad \qquad 0 \\
0 & 0 & -\sinh^{-2}{\rho}\,\cosh^{-2}{\rho}
\end{array} \right)_{\mu\nu}.
\label{grumillermetric}
\end{aligned}
\end{equation} 
Considering the Killing vector $\bar{\xi} = (-1,0,0)$ one finds the result of the integral in (\ref{ttt}) to be  non-vanishing 
\begin{equation}
Q_{Taub}\left[\bar{\text{\ensuremath{\xi}}}\right] = \frac{\pi}{2 \ell} \bigg( 3 \tau^2 - \frac{ 161}{72} \bigg),
\end{equation}
which shows that this log mode is not in the tangent space of the solution space of chiral gravity around the $AdS_3$ metric. 

\section{Conclusions and Discussions}

We have shown that at certain {\it critical parameter } values of extended gravity theories in constant curvature backgrounds, perturbation theory fails. Our arguments provide  support to the discussion given by \cite{Strom2, Carlip} regarding the linearization instability in three dimensional chiral gravity and extend the discussion to generic gravity theories in a somewhat rigorous form. The crucial point is that even in spacetimes with non-compact Cauchy surfaces, linearization instability can exist for background metrics with at least one Killing vector field.  Our computation also sheds light on the earlier observations \cite{Deser_Tekin} that at certain critical values of the parameters defining the theory, conserved charges of all solutions, such as black holes, excitations  vanish identically.\footnote{For a recent review of conserved charges in generic gravity theories see the book \cite{kitap}.} For example,  Kerr-AdS black hole metrics  have the same mass and angular momentum as the AdS background.  This leads to a rather non-physical infinite degeneracy of the vacuum: for example, creating back holes costs nothing which is unacceptable. With our discussion above, it is now clear that,  perturbation theory which is used to define boundary integrals of the conserved Killing charges does not make sense exactly at the critical values of the parameters. Therefore  one really needs a new method to find/define conserved charges in these  theories at their critical points. One such method was proposed in  for quadratic theories \cite{Deser_Tekin_2007} and in \cite{sezginefendi} for TMG. 

We must note that, for asymptotically flat spacetimes, the ADM mass is the correct definition of mass-energy for any metric-based theory of gravity. Therefore, the stability of the Minkowski space as was shown for Einstein's theory by Choquet-Bruhat and Deser \cite{Choquet-Bruhat}  is valid for all higher derivative models as long as one considers the non-compact  Cauchy surfaces and asymptotically flat boundary conditions.  But once a cosmological  constant is introduced, the problem changes dramatically as we have shown: the ADM mass-energy (or angular momentum) expressions are modified and conserved charges get contributions from each covariant  tensors added to the field equations. Once such a construction is understood, it is clear that 
some theories will have identically vanishing charges for all solutions with some fixed boundary conditions, which is a signal of linearization instability.  

It is also important to realize that, linearization instability of certain background solutions in some theories is not bad as it sounds: for example chiral gravity is a candidate   both as a non-trivial classical and quantum gravity theory in $AdS_3$ with a two dimensional chiral conformal field theory induced on the boundary. But it has log-mode solutions which appear as ghosts in the classical theory and negative norm states in the quantum theory.  It just turns out that chiral gravity in $AdS_3$ has linearization instability along these log-modes: namely, they do not have vanishing Taub conserved quantities which is a constraint for all integrable solutions. 
Therefore, they cannot come from linearization of exact solutions.  A similar phenomenon takes place for the minimal massive gravity \cite{Bergshoeff} which was proposed as a possible solution to the bulk-boundary unitarity clash in three dimensional gravity theories and as a viable model that has a healthy dual conformal field theory on the boundary of $AdS_3$. It was shown recently in \cite{Ercan} that this theory  only makes sense at the chiral point \cite{Tekin, mom} and hence linearization instability arises at that point which can save the theory from its log-modes. Let us note that we have also computed the second order constraint in the minimal massive gravity, namely the Taub conserved quantity and found that it is non-vanishing.

In the discussion of linearization stability and instability of a given exact solution in the context of general relativity, we noted that to make use of the powerful techniques of elliptic operator theory, one rewrites the 
four dimensional Einstein's theory as a dynamical system with constraints on a spacelike Cauchy surface and  
the evolution equations.  As the constraints are intact, initial Cauchy data uniquely defines a spacetime 
(modulo some technical assumptions). Therefore, to study the linearization stability one can simply study the linearization stability of the constraints on the surface where the metric  tensor field is positive-definite. 
All these arguments boil down to showing that the initial background metric is not a singular point and that the space of solutions around the initial metric is an open subset (in fact a submanifold) of all solutions. This can be shown by proving the surjectivity of the operators that appear in the linearized constraints. A similar construction, dynamical formulation of the  higher derivative models studied here in AdS and the surjectivity of the relevant linear maps would be highly valuable.  
For the case of the cosmological Einstein's theory, such a construction was carried out in \cite{2009_hyperbolic} where it was observed that certain strong decays lead to linearization instability even for non-compact Cauchy surfaces with hyperbolic asymptotics.

\section*{Appendix A: Second order perturbation theory and gauge invariance issues }

Here without going into too much detail let us summarize some of the relevant formulas that we use in the bulk of the paper to show various expressions, such as the gauge transformation of the background tensors, second order forms of the tensors {\it etc.}

Lie and covariant derivatives do not commute so we shall need the following expressions. Let $X$ be a vector field on our manifold with a metric $\bar g$ and $T$ be a $(0,2)$ background tensor field. Then in components one has the Lie derivative of $T$ with respect to $X$ as
\begin{equation}
\text{\ensuremath{\mathscr{L}}}_{X}T_{\rho\sigma}=X^{f}\bar{\nabla}_{f}T_{\rho\sigma}+\left(\bar{\nabla}_{\rho}X^{f}\right)T_{f\sigma}+\left(\bar{\nabla}_{\sigma}X^{f}\right)T_{\rho f}.
\end{equation}
Then one has the following difference of the derivatives
\begin{equation}
\bar{\nabla}_{\mu}\text{\ensuremath{\mathscr{L}}}_{X}T_{\rho\sigma}-\text{\ensuremath{\mathscr{L}}}_{X}\bar{\nabla}_{\mu}T_{\rho\sigma}=\left(\bar{\nabla}_{\mu}\bar{\nabla}_{\rho}X^{f}+X^{\lambda}\bar{R}_{\mu\lambda\rho}\thinspace^{f}\right)T_{f\sigma}+\left(\bar{\nabla}_{\mu}\bar{\nabla}_{\sigma}X^{f}+X^{\lambda}\bar{R}_{\mu\lambda\sigma}\thinspace^{f}\right)T_{f\rho}.
\end{equation}
Let $\delta_{X}$ denote the gauge transformation generated by $X$, then the gauge transformation of the Christoffel connection reads,
\begin{equation}
\delta_{X}(\Gamma_{\mu\nu}\thinspace^{\gamma})^{(1)}=\bar{\nabla}_{\mu}\bar{\nabla}_{\nu}X^{\gamma}+\bar{R}^{\gamma}\thinspace_{\nu\sigma\mu}X^{\sigma}.
\end{equation}
Making use of this one finds
\begin{equation}
\bar{\nabla}_{\mu}\text{\ensuremath{\mathscr{L}}}_{X}T_{\rho\sigma}=\text{\ensuremath{\mathscr{L}}}_{X}\bar{\nabla}_{\mu}T_{\rho\sigma}+T_{\alpha\sigma}\delta_{X}(\Gamma_{\mu\rho}\thinspace^{\alpha})^{(1)}+T_{\rho\alpha}\delta_{X}(\Gamma_{\mu\sigma}\thinspace^{\alpha})^{(1)}.
\label{id1}
\end{equation}
Applying the same procedure for the case of any generic three index
tensor, we arrive the relation
\begin{equation}
\bar{\nabla}_{\mu}\text{\ensuremath{\mathscr{L}}}_{X}T_{\rho\sigma\gamma}=\text{\ensuremath{\mathscr{L}}}_{X}\bar{\nabla}_{\mu}T_{\rho\sigma\gamma}+T_{\alpha\sigma\gamma}\delta_{X}(\Gamma_{\mu\rho}\thinspace^{\alpha})^{(1)}+T_{\rho\alpha\gamma}\delta_{X}(\Gamma_{\mu\sigma}\thinspace^{\alpha})^{(1)}+T_{\rho\sigma\alpha}\delta_{X}(\Gamma_{\mu\gamma}\thinspace^{\alpha})^{(1)}.
\label{id2}
\end{equation}

Let us summarize some results about the second order perturbation theory (see also \cite{Tahsin_born}).
By definition one has 
\begin{equation}
g_{\mu\nu}:=\bar{g}_{\mu\nu}+\tau h_{\mu\nu},
\end{equation}
whose inverse is
\begin{equation}
g^{\mu\nu}=\bar{g}^{\mu\nu}+\tau h^{\mu\nu}+\tau^{2}h_{\alpha}^{\mu}h^{\alpha\nu}+O(\tau^{3}).
\end{equation}
Let  $T$ be a generic tensor, then it can be expanded as 
\begin{equation}
T=\bar T+\tau {T}^{(1)}+\tau^{2}{{T}}^{(2)} +O(\tau^{3}).
\end{equation}
For the Christoffel connection we have
\begin{equation}
\varGamma_{\mu\nu}\thinspace^{\gamma}=\bar{\varGamma}_{\mu\nu}\thinspace^{\gamma}+\tau(\varGamma_{\mu\nu}\thinspace^{\gamma})^{(1)}+\tau^{2}(\varGamma_{\mu\nu}\thinspace^{\gamma})^{(2)},
\end{equation}
where the first order term is 
\begin{equation}
(\Gamma_{\mu\nu}\thinspace^{\gamma})^{(1)}=\frac{1}{2}\big (\bar{\nabla}_{\mu}h_{\nu}^{\gamma}+\bar{\nabla}_{\nu}h_{\mu}^{\gamma}-\bar{\nabla}^{\gamma}h_{\mu\nu}),
\end{equation}
and the second order one is
\begin{equation}
(\Gamma_{\mu\nu}\thinspace^{\gamma})^{(2)}=-h^{\gamma\delta}(\Gamma_{\mu\nu\delta})^{(1)}.
\end{equation}
Since it is a background tensor, we can raise and lower the indices with $\bar{g}_{\mu\nu}$
\begin{equation}
(\Gamma_{\mu\nu\delta})^{(1)}=\bar{g}_{\gamma\delta}(\Gamma_{\mu\nu}\thinspace^{\gamma})^{(2)}.
\end{equation}
The first order linearized Riemann tensor is
\begin{equation}
(R^{\rho}\thinspace_{\mu\sigma\nu})^{(1)}=\bar{\nabla}_{\sigma}(\Gamma_{\nu\mu}\thinspace^{\rho})^{(1)}-\bar{\nabla}_{\nu}(\Gamma_{\sigma\mu}\thinspace^{\rho})^{(1)},
\end{equation}
and the second order linearized Riemann tensor is
\begin{equation}
(R^{\rho}\thinspace_{\mu\sigma\nu})^{(2)}=\bar{\nabla}_{\sigma}(\Gamma_{\nu\mu}\thinspace^{\rho})^{(2)}-\bar{\nabla}_{\nu}(\Gamma_{\sigma\mu}\thinspace^{\rho})^{(2)}+(\Gamma_{\mu\nu}\thinspace^{\alpha})^{(1)}(\Gamma_{\sigma\alpha}\thinspace^{\rho})^{(1)}-(\Gamma_{\mu\sigma}\thinspace^{\alpha})^{(1)}(\Gamma_{\nu\alpha}\thinspace^{\rho})^{(1)}.
\end{equation}
The first order linearized Ricci tensor is
\begin{equation}
(R_{\mu\nu})^{(1)}=\bar{\nabla}_{\sigma}(\Gamma_{\mu\nu}\thinspace^{\sigma})^{(1)}-\bar{\nabla}_{\nu}(\Gamma_{\sigma\mu}\thinspace^{\sigma})^{(1)},
\end{equation}
and the second order linearized Ricci tensor is
\begin{equation}
(R_{\mu\nu})^{(2)}=\bar{\nabla}_{\sigma}(\Gamma_{\nu\mu}\thinspace^{\sigma})^{(2)}-\bar{\nabla}_{\nu}(\Gamma_{\sigma\mu}\thinspace^{\sigma})^{(2)}+(\Gamma_{\mu\nu}\thinspace^{\alpha})^{(1)}(\Gamma_{\sigma\alpha}\thinspace^{\sigma})^{(1)}-(\Gamma_{\mu\sigma}\thinspace^{\alpha})^{(1)}(\Gamma_{\nu\alpha}\thinspace^{\sigma})^{(1)}.
\end{equation}
We shall need the explicit form of it in terms of the $h_{\mu\nu}$ field which reads
\begin{multline*}
(R_{\mu\nu})^{(2)}=-\frac{1}{2}\bar{\nabla}_{\rho}\left[h^{\rho\beta}(\bar{\nabla}_{\mu}h_{\nu\beta}+\bar{\nabla}_{\nu}h_{\mu\beta}-\bar{\nabla}_{\beta}h_{\nu\mu})\right]+\frac{1}{2}\bar{\nabla}_{\nu}\left[h^{\rho\beta}\bar{\nabla}_{\mu}h_{\rho\beta}\right]-\frac{1}{4}\left(\bar{\nabla}_{\mu}h_{\rho\beta}\right)\bar{\nabla}_{\nu}h^{\rho\beta}\\
+\frac{1}{4}\left(\bar{\nabla}^{\beta}h\right)(\bar{\nabla}_{\mu}h_{\nu\beta}+\bar{\nabla}_{\nu}h_{\mu\beta}-\bar{\nabla}_{\beta}h_{\nu\mu})+\frac{1}{2}(\bar{\nabla}_{\beta}h_{\nu\alpha})\bar{\nabla}^{\beta}h_{\mu}^{\alpha}-\frac{1}{2}(\bar{\nabla}_{\beta}h_{\nu\alpha})\bar{\nabla}^{\alpha}h_{\mu}^{\beta}.
\end{multline*}
The linearized scalar curvature is
\begin{equation}
(R)^{(1)}=\bar{\nabla}_{\alpha}\bar{\nabla}_{\beta}h^{\alpha\beta}-\bar{\text{\textifsymbol[ifgeo]{48}}}h-\bar{R}_{\mu\nu}h^{\mu\nu},
\end{equation}
and the second order linearized scalar curvature is
\begin{equation}
(R)^{(2)}=\bar{R}_{\mu\nu}h_{\alpha}^{\mu}h^{\alpha\nu}-(R_{\mu\nu})^{(1)}h^{\mu\nu}+\bar{g}^{\mu\nu}(R_{\mu\nu})^{(2)}.
\end{equation}
Explicitly we have
\begin{multline*}
(R)^{(2)}=-\frac{1}{2}\bar{\nabla}_{\rho}\left[h^{\rho\beta}(2\bar{\nabla}_{\sigma}h_{\beta}^{\sigma}-\bar{\nabla}_{\beta}h)\right]+\frac{1}{2}\bar{\nabla}_{\sigma}\left[h^{\rho\beta}\bar{\nabla}^{\sigma}h_{\rho\beta}\right]+\frac{1}{4}\left(\bar{\nabla}^{\beta}h\right)(2\bar{\nabla}_{\sigma}h_{\beta}^{\sigma}-\bar{\nabla}_{\beta}h)\\
+\frac{1}{4}\left(\bar{\nabla}_{\sigma}h_{\rho\beta}\right)\bar{\nabla}^{\sigma}h^{\rho\beta}-\frac{1}{2}\left(\bar{\nabla}_{\sigma}h_{\rho\beta}\right)\bar{\nabla}^{\rho}h^{\sigma\beta}-\frac{1}{2}h^{\rho\beta}\left[2\bar{\nabla}_{\sigma}\bar{\nabla}_{\rho}h_{\beta}^{\sigma}-\bar{\text{\textifsymbol[ifgeo]{48}}}h_{\rho\beta}-\bar{\nabla}_{\rho}\bar{\nabla}_{\beta}h\right]+\bar{R}_{\rho\beta}h^{\rho\alpha}h_{\alpha}^{\beta}.
\end{multline*}
Using the above results, let us find how the second order linearized form of the Einstein tensor transforms under the gauge transformations generated by the flow of $X$. In the index-free notation one has
\begin{equation}
D^{2}Ein(^{(4)}g)\cdot\left(^{(4)}h,\text{\ensuremath{\mathscr{L}}}_{(4)X}{}^{(4)}g\right)+DEin(^{(4)}g)\cdot\text{\ensuremath{\mathscr{L}}}_{(4)X}{}^{(4)}h=\text{\ensuremath{\mathscr{L}}}_{(4)X}\left(DEin(^{(4)}g)\cdot{}^{(4)}h\right),
\end{equation}
which reads in local coordinates as
\begin{equation}
\delta_{X}(G_{\mu\nu})^{(2)}\cdot [h,h]+(G_{\mu\nu})^{(1)}\cdot\text{\ensuremath{\mathscr{L}}}_{X}h=\text{\ensuremath{\mathscr{L}}}_{X}(G_{\mu\nu})^{(1)}\cdot h.
\end{equation}
Let us prove this. By definition we have
\begin{equation}
\delta_{X}(G_{\mu\nu})^{(2)}\cdot [h,h]=\delta_{X}(R_{\mu\nu})^{(2)}\cdot [h,h]-\frac{1}{2}\bar{g}_{\mu\nu}\delta_{X}(R)^{(2)}\cdot [h,h]-\frac{1}{2}(R)^{(1)}\cdot h\delta_{X}h_{\mu\nu}-\frac{1}{2}h_{\mu\nu}\delta_{X}(R)^{(1)}\cdot h.
\end{equation}
Let us calculate the right hand side of the equation term by term.
\begin{multline*}
\delta_{X}(R{}_{\mu\nu})^{(2)}\cdot [h,h]=-\left(\delta_{X}h^{\rho\beta}\right)\left(\bar{\nabla}_{\rho}(\Gamma_{\nu\mu\beta})^{(1)}-\bar{\nabla}_{\nu}(\Gamma_{\rho\mu\beta})^{(1)}\right)-h^{\rho\beta}\delta_{X}\left(\bar{\nabla}_{\rho}(\Gamma_{\nu\mu\beta})^{(1)}-\bar{\nabla}_{\nu}(\Gamma_{\rho\mu\beta})^{(1)}\right)\\
-\delta_{X}\left((\Gamma_{\mu\nu}\thinspace^{\alpha})^{(1)}(\Gamma_{\sigma}\thinspace^{\sigma}\thinspace_{\alpha})^{(1)}-(\Gamma_{\mu\sigma}\thinspace^{\alpha})^{(1)}(\Gamma_{\nu}\thinspace^{\sigma}\thinspace_{\alpha})^{(1)}\right).
\end{multline*}
Since one has 
\begin{equation}
\delta_{X}h^{\rho\beta}=-\text{\ensuremath{\mathscr{L}}}_{X}\bar{g}^{\rho\beta},
\end{equation}
using the identities (\ref{id1}, \ref{id2}) we have
\begin{multline*}
\delta_{X}(R{}_{\mu\nu})^{(2)}\cdot [h,h]=\text{\ensuremath{\mathscr{L}}}_{X}(R_{\mu\nu})^{(1)}\cdot h-\frac{1}{2}\bar{g}^{\rho\beta}\left[\bar{\nabla}_{\rho}\text{\ensuremath{\mathscr{L}}}_{X}\left(\bar{\nabla}_{\nu}h_{\mu\beta}+\bar{\nabla}_{\mu}h_{\nu\beta}-\bar{\nabla}_{\beta}h_{\mu\nu}\right)-\bar{\nabla}_{\nu}\text{\ensuremath{\mathscr{L}}}_{X}\bar{\nabla}_{\mu}h_{\rho\beta}\right]\\
+\left(\bar{\nabla}_{\nu}h_{\sigma}\thinspace^{\rho}\right)\delta_{X}(\Gamma_{\rho\mu}\thinspace^{\sigma})^{(1)}-\left(\bar{\nabla}_{\rho}h_{\sigma}\thinspace^{\rho}\right)\delta_{X}(\Gamma_{\nu\mu}\thinspace^{\sigma})^{(1)}-h^{\rho}\thinspace_{\beta}\delta_{X}\left(R^{\beta}\thinspace_{\mu\rho\nu}\right)^{(1)}\cdot h.
\end{multline*}
Finally one can find 
\begin{equation}
\delta_{X}(R{}_{\mu\nu})^{(2)}\cdot [h,h]=\text{\ensuremath{\mathscr{L}}}_{X}(R_{\mu\nu})^{(1)}\cdot h-\left(R_{\mu\nu}\right)^{(1)}\cdot\text{\ensuremath{\mathscr{L}}}_{X}h,
\end{equation}
and from the following definition
\begin{equation}
\left(R\right)^{(2)}\cdot [h,h]=\bar{R}_{\rho\sigma}h^{\sigma\lambda}h_{\lambda}\thinspace^{\rho}-h^{\sigma\rho}(R_{\rho\sigma})^{(1)}\cdot h+\bar{g}^{\sigma\lambda}(R_{\rho\sigma})^{(2)}\cdot [h,h],
\end{equation}
 one can find 
\begin{equation}
\delta_{X}\left(R\right)^{(2)}\cdot [h,h]=\text{\ensuremath{\mathscr{L}}}_{X}\left(R\right)^{(1)}\cdot h-\left[\bar{g}^{\sigma\rho}\left(R_{\rho\sigma}\right)^{(1)}\cdot\text{\ensuremath{\mathscr{L}}}_{X}h-\bar{R}^{\rho\sigma}\text{\ensuremath{\mathscr{L}}}_{X}h_{\sigma\rho}\right],
\end{equation}
which can be reduced to 
\begin{equation}
\delta_{X}\left(R\right)^{(2)}\cdot [h,h]=\text{\ensuremath{\mathscr{L}}}_{X}\left(R\right)^{(1)}\cdot h-\left(R\right)^{(1)}\cdot\text{\ensuremath{\mathscr{L}}}_{X}h.
\end{equation}
Then we can collect these to get the gauge transformation of the second order expansion of the Einstein tensor as  
\begin{multline*}
\delta_{X}(G_{\mu\nu})^{(2)}\cdot [h,h]=\text{\ensuremath{\mathscr{L}}}_{X}\left[(R_{\mu\nu})^{(1)}\cdot h-\frac{1}{2}\bar{g}_{\mu\nu}\left(R\right)^{(1)}\cdot h-\frac{1}{2}h_{\mu\nu}\bar{R}\right]\\
-\left[\left(R_{\mu\nu}\right)^{(1)}\cdot\text{\ensuremath{\mathscr{L}}}_{X}h-\frac{1}{2}\bar{g}_{\mu\nu}\left(R\right)^{(1)}\cdot\text{\ensuremath{\mathscr{L}}}_{X}h-\frac{1}{2}\bar{R}\text{\ensuremath{\mathscr{L}}}_{X}h_{\mu\nu}\right].\\
\end{multline*}
The first line is the Lie derivative of the linearized Einstein tensor
and the second line is the linearized Einstein tensor evaluated at $\text{\ensuremath{\mathscr{L}}}_{X}h$.
\begin{equation}
\delta_{X}(G_{\mu\nu})^{(2)}\cdot [h,h]=\text{\ensuremath{\mathscr{L}}}_{X}(G_{\mu\nu})^{(1)}\cdot h-\left(G_{\mu\nu}\right)^{(1)}\cdot\text{\ensuremath{\mathscr{L}}}_{X}h,
\end{equation}
which is the desired formula. 

\section*{Appendix B: Explicit form of the $K_{\mu \nu}$ tensor in AdS }

Here let us depict some of the intermediate steps leading to (\ref{div1}). Assuming a general form for the $k_{\mu \nu}$ as 
\begin{equation}
k_{\mu\nu}=ah_{\mu\beta}h_{\nu}^{\beta}+bhh_{\mu\nu}+\bar{g}_{\mu\nu}(ch_{\alpha\beta}^{2}+dh^{2}),
\end{equation}
the first order Ricci  operator evaluated at $k$ is 
\begin{equation}
(R_{\mu\nu})^{(1)}\cdot k=\frac{1}{2}(\bar{\nabla}_{\alpha}\bar{\nabla}_{\mu}k_{\nu}^{\alpha}+\bar{\nabla}_{\alpha}\bar{\nabla}_{\nu}k_{\mu}^{\alpha}-\bar{\text{\textifsymbol[ifgeo]{48}}}k_{\mu\nu}-\bar{\nabla}_{\mu}\bar{\nabla}_{\nu}k),
\end{equation}
whose explicit form follows as 
\begin{multline*}
(R_{\mu\nu})^{(1)}\cdot k=\frac{a}{2}(\bar{\nabla}_{\alpha}\bar{\nabla}_{\mu}h^{\alpha\beta}h_{\beta\nu}+\bar{\nabla}_{\alpha}\bar{\nabla}_{\nu}h^{\alpha\beta}h_{\beta\mu}-\bar{\text{\textifsymbol[ifgeo]{48}}}h_{\nu}^{\beta}h_{\beta\mu}-\bar{\nabla}_{\mu}\bar{\nabla}_{\nu}h_{\alpha\beta}^{2})\\
+\frac{b}{2}(\bar{\nabla}_{\alpha}\bar{\nabla}_{\mu}hh_{\nu}^{\alpha}+\bar{\nabla}_{\alpha}\bar{\nabla}_{\nu}hh_{\mu}^{\alpha}-\bar{\text{\textifsymbol[ifgeo]{48}}}hh_{\mu\nu}-\bar{\nabla}_{\mu}\bar{\nabla}_{\nu}h^{2})\\
+\frac{c}{2}(\bar{\nabla}_{\nu}\bar{\nabla}_{\mu}h_{\alpha\beta}^{2}+\bar{\nabla}_{\mu}\bar{\nabla}_{\nu}h_{\alpha\beta}^{2}-\bar{g}_{\mu\nu}\bar{\text{\textifsymbol[ifgeo]{48}}}h_{\alpha\beta}^{2}-D\bar{\nabla}_{\mu}\bar{\nabla}_{\nu}h_{\alpha\beta}^{2})\\
+\frac{d}{2}(\bar{\nabla}_{\nu}\bar{\nabla}_{\mu}h^{2}+\bar{\nabla}_{\mu}\bar{\nabla}_{\nu}h^{2}-\bar{g}_{\mu\nu}\bar{\text{\textifsymbol[ifgeo]{48}}}h^{2}-D\bar{\nabla}_{\mu}\bar{\nabla}_{\nu}h^{2}).\thinspace\thinspace\thinspace\thinspace\thinspace\thinspace\thinspace\thinspace\thinspace\thinspace\thinspace\thinspace\thinspace\thinspace\thinspace\thinspace\thinspace\thinspace\thinspace\thinspace\thinspace\thinspace\thinspace\thinspace\thinspace\thinspace\thinspace\thinspace\thinspace\thinspace\thinspace\thinspace\thinspace\thinspace\thinspace\thinspace\thinspace\thinspace\thinspace\thinspace\thinspace\thinspace\thinspace\thinspace\thinspace\thinspace\thinspace\thinspace\thinspace\thinspace
\end{multline*}
We should set $a=1$ and $b=-1/2$ to get the second order linearized Ricci tensor
\begin{eqnarray}
(R_{\mu\nu})^{(1)}\cdot k=\frac{1}{2}\bar{\nabla}_{\alpha}\left(h^{\alpha\beta}\left(\bar{\nabla}_{\mu}h_{\nu\beta}+\bar{\nabla}_{\nu}h_{\mu\beta}\right)\right)-\frac{1}{4}\bar{\nabla}^{\beta}h\left(\bar{\nabla}_{\mu}h_{\nu\beta}+\bar{\nabla}_{\nu}h_{\mu\beta}-\bar{\nabla}_{\beta}h_{\mu\nu}\right)\nonumber\\
+\frac{1}{2}\bar{\nabla}_{\alpha}\left(h_{\beta\nu}\bar{\nabla}_{\mu}h^{\alpha\beta}+h_{\beta\mu}\bar{\nabla}_{\nu}h^{\alpha\beta}+h^{\alpha\beta}\bar{\nabla}_{\beta}h_{\nu\mu}-\bar{\nabla}^{\alpha}\left(h_{\nu}^{\beta}h_{\beta\mu}\right)\right)\nonumber\\
-\bar{\nabla}_{\nu}\left(h^{\alpha\beta}\bar{\nabla}_{\mu}h_{\alpha\beta}\right)-\frac{1}{4}h\bar{\nabla}_{\alpha}\left(\bar{\nabla}_{\mu}h_{\nu}^{\alpha}+\bar{\nabla}_{\nu}h_{\mu}^{\alpha}-\bar{\nabla}^{\alpha}h_{\mu\nu}\right)+\frac{1}{2}\bar{\nabla}_{\nu}\left(h\bar{\nabla}_{\mu}h\right)\nonumber\\
-\frac{1}{4}\bar{\nabla}_{\alpha}(h_{\nu}^{\alpha}\bar{\nabla}_{\mu}h+h_{\mu}^{\alpha}\bar{\nabla}_{\nu}h-h_{\mu\nu}\bar{\nabla}^{\alpha}h)\nonumber\\
+\frac{c}{2}((2-D)\bar{\nabla}_{\nu}\bar{\nabla}_{\mu}h_{\alpha\beta}^{2}-\bar{g}_{\mu\nu}{\bar\square}h_{\alpha\beta}^{2})+\frac{d}{2}((2-D)\bar{\nabla}_{\nu}\bar{\nabla}_{\mu}h^{2}-\bar{g}_{\mu\nu}\bar{\bar\square}h^{2}).
\end{eqnarray}

Finally the Ricci tensor evaluated at $k$ becomes
\begin{multline*}
(R_{\mu\nu})^{(1)}\cdot k=-(R_{\mu\nu})^{(2)}\cdot [h,h]-\frac{3}{4}\bar{\nabla}_{\nu}h^{\alpha\beta}\bar{\nabla}_{\mu}h_{\alpha\beta}+\frac{1}{2}\bar{\nabla}_{\alpha}h_{\mu\beta}\bar{\nabla}^{\alpha}h_{\nu}^{\beta}-\frac{1}{2}\bar{\nabla}_{\alpha}h_{\mu\beta}\bar{\nabla}^{\beta}h_{\nu}^{\alpha}\\
+\frac{1}{2}\bar{\nabla}_{\alpha}\left(h_{\beta\nu}\bar{\nabla}_{\mu}h^{\alpha\beta}+h_{\beta\mu}\bar{\nabla}_{\nu}h^{\alpha\beta}+h^{\alpha\beta}\bar{\nabla}_{\beta}h_{\nu\mu}-\bar{\nabla}^{\alpha}\left(h_{\nu}^{\beta}h_{\beta\mu}\right)\right)\\
-\frac{1}{2}h^{\alpha\beta}\bar{\nabla}_{\nu}\bar{\nabla}_{\mu}h_{\alpha\beta}-\frac{h}{2}(R_{\mu\nu})^{(1)}\cdot h-\frac{1}{4}h\bar{\nabla}_{\nu}\bar{\nabla}_{\mu}h+\frac{1}{2}\bar{\nabla}_{\nu}\left(h\bar{\nabla}_{\mu}h\right)-\frac{1}{4}\bar{\nabla}_{\alpha}(h_{\nu}^{\alpha}\bar{\nabla}_{\mu}h+h_{\mu}^{\alpha}\bar{\nabla}_{\nu}h-h_{\mu\nu}\bar{\nabla}^{\alpha}h)\\
+\frac{c}{2}((2-D)\bar{\nabla}_{\nu}\bar{\nabla}_{\mu}h_{\alpha\beta}^{2}-\bar{g}_{\mu\nu}\bar{\text{\textifsymbol[ifgeo]{48}}}h_{\alpha\beta}^{2})+\frac{d}{2}((2-D)\bar{\nabla}_{\nu}\bar{\nabla}_{\mu}h^{2}-\bar{g}_{\mu\nu}\bar{\text{\textifsymbol[ifgeo]{48}}}h^{2}),\thinspace\thinspace\thinspace\thinspace\thinspace\thinspace\thinspace\thinspace\thinspace\thinspace\thinspace\thinspace\thinspace\thinspace\thinspace\thinspace\thinspace\thinspace\thinspace\thinspace\thinspace
\end{multline*}
from which one can find  the scalar curvature
\begin{multline*}
(R)^{(1)}\cdot k=-(R)^{(2)}\cdot [ h,h]-\frac{5}{4}\bar{\nabla}^{\mu}h^{\alpha\beta}\bar{\nabla}_{\mu}h_{\alpha\beta}+\frac{1}{2}\bar{\nabla}_{\alpha}h_{\mu\beta}\bar{\nabla}^{\beta}h^{\mu\alpha}\\
+\frac{1}{2}h^{\alpha\beta}\bar{\nabla}_{\alpha}\bar{\nabla}_{\beta}h-h^{\alpha\beta}\bar{\text{\textifsymbol[ifgeo]{48}}}h_{\alpha\beta}-\frac{h}{2}(R)^{(1)}\cdot h+\frac{1}{2}h\bar{\text{\textifsymbol[ifgeo]{48}}}h+\frac{3}{4}\bar{\nabla}^{\mu}h\bar{\nabla}_{\mu}h\\
+c(1-D)\bar{\text{\textifsymbol[ifgeo]{48}}}h_{\alpha\beta}^{2}+d(1-D)\bar{\text{\textifsymbol[ifgeo]{48}}}h^{2}-\bar{R}(ch_{\alpha\beta}^{2}+dh^{2}).
\end{multline*}
Finally the linearized Einstein tensor can be found as 
\begin{multline*}
\left(\text{\ensuremath{\mathcal{G}}}_{\mu\nu}\right)^{(1)}\cdot k=-\left(\text{\ensuremath{\mathcal{G}}}_{\mu\nu}\right)^{(2)}\cdot [ h,h]-\frac{1}{2}h_{\mu\nu}\left({R}\right)^{(1)}\cdot h-\frac{h}{2}(\mathcal{G}{}_{\mu\nu})^{(1)}\cdot h\\
-\frac{3}{4}\bar{\nabla}_{\nu}h^{\alpha\beta}\bar{\nabla}_{\mu}h_{\alpha\beta}+\frac{1}{2}\bar{\nabla}_{\alpha}h_{\mu\beta}\bar{\nabla}^{\alpha}h_{\nu}^{\beta}-\frac{1}{2}\bar{\nabla}_{\alpha}h_{\mu\beta}\bar{\nabla}^{\beta}h_{\nu}^{\alpha}\\
+\frac{1}{2}\bar{\nabla}_{\alpha}\left(h_{\beta\nu}\bar{\nabla}_{\mu}h^{\alpha\beta}+h_{\beta\mu}\bar{\nabla}_{\nu}h^{\alpha\beta}+h^{\alpha\beta}\bar{\nabla}_{\beta}h_{\nu\mu}-\bar{\nabla}^{\alpha}\left(h_{\nu}^{\beta}h_{\beta\mu}\right)\right)\\
-\frac{1}{2}h^{\alpha\beta}\bar{\nabla}_{\nu}\bar{\nabla}_{\mu}h_{\alpha\beta}-\frac{1}{4}h\bar{\nabla}_{\nu}\bar{\nabla}_{\mu}h+\frac{1}{2}\bar{\nabla}_{\nu}\left(h\bar{\nabla}_{\mu}h\right)\\
-\frac{1}{4}\bar{\nabla}_{\alpha}(h_{\nu}^{\alpha}\bar{\nabla}_{\mu}h+h_{\mu}^{\alpha}\bar{\nabla}_{\nu}h-h_{\mu\nu}\bar{\nabla}^{\alpha}h)+\frac{c}{2}(2-D)\bar{\nabla}_{\nu}\bar{\nabla}_{\mu}h_{\alpha\beta}^{2}+\frac{d}{2}(2-D)\bar{\nabla}_{\nu}\bar{\nabla}_{\mu}h^{2}\\
-\frac{1}{2}\bar{g}_{\mu\nu}\left[-\frac{5}{4}\bar{\nabla}^{\sigma}h^{\alpha\beta}\bar{\nabla}_{\sigma}h_{\alpha\beta}+\frac{1}{2}\bar{\nabla}_{\alpha}h_{\sigma\beta}\bar{\nabla}^{\beta}h^{\sigma\alpha}\right]\\
-\frac{1}{2}\bar{g}_{\mu\nu}\left[\frac{1}{2}h^{\alpha\beta}\bar{\nabla}_{\alpha}\bar{\nabla}_{\beta}h-h^{\alpha\beta}\bar{\text{\textifsymbol[ifgeo]{48}}}h_{\alpha\beta}+\frac{1}{2}h\bar{\text{\textifsymbol[ifgeo]{48}}}h+\frac{3}{4}\bar{\nabla}^{\sigma}h\bar{\nabla}_{\sigma}h\right]\\
-\frac{1}{2}\bar{g}_{\mu\nu}\left[c(2-D)\bar{\text{\textifsymbol[ifgeo]{48}}}h_{\alpha\beta}^{2}+d(2-D)\bar{\text{\textifsymbol[ifgeo]{48}}}h^{2})-\bar{R}(ch_{\alpha\beta}^{2}+dh^{2})\right]-\frac{2\Lambda}{D-2}\left[h_{\mu\beta}h_{\nu}^{\beta}+\bar{g}_{\mu\nu}(ch_{\alpha\beta}^{2}+dh^{2})\right].
\end{multline*}
Using these, one can find the final form of the $K_{\mu \nu}$ tensor as 
\begin{multline*}
K_{\mu\nu}=-\frac{1}{2}h_{\mu\nu}\left({R}\right)^{(1)}\cdot h-\frac{h}{2}(\text{\ensuremath{\mathcal{G}}}{}_{\mu\nu})^{(1)}\cdot h-\frac{3}{4}\bar{\nabla}_{\nu}h^{\alpha\beta}\bar{\nabla}_{\mu}h_{\alpha\beta}+\frac{1}{2}\bar{\nabla}_{\alpha}h_{\mu\beta}\bar{\nabla}^{\alpha}h_{\nu}^{\beta}\\
-\frac{1}{2}\bar{\nabla}_{\alpha}h_{\mu\beta}\bar{\nabla}^{\beta}h_{\nu}^{\alpha}+\frac{1}{2}\bar{\nabla}_{\alpha}\left(h_{\beta\nu}\bar{\nabla}_{\mu}h^{\alpha\beta}+h_{\beta\mu}\bar{\nabla}_{\nu}h^{\alpha\beta}+h^{\alpha\beta}\bar{\nabla}_{\beta}h_{\nu\mu}-\bar{\nabla}^{\alpha}\left(h_{\nu}^{\beta}h_{\beta\mu}\right)\right)\\
-\frac{1}{2}h^{\alpha\beta}\bar{\nabla}_{\nu}\bar{\nabla}_{\mu}h_{\alpha\beta}-\frac{1}{4}h\bar{\nabla}_{\nu}\bar{\nabla}_{\mu}h+\frac{1}{2}\bar{\nabla}_{\nu}\left(h\bar{\nabla}_{\mu}h\right)-\frac{1}{4}\bar{\nabla}_{\alpha}(h_{\nu}^{\alpha}\bar{\nabla}_{\mu}h+h_{\mu}^{\alpha}\bar{\nabla}_{\nu}h-h_{\mu\nu}\bar{\nabla}^{\alpha}h)\\
+\frac{c}{2}(2-D)\bar{\nabla}_{\nu}\bar{\nabla}_{\mu}h_{\alpha\beta}^{2}+\frac{d}{2}(2-D)\bar{\nabla}_{\nu}\bar{\nabla}_{\mu}h^{2}-\frac{2\Lambda}{D-2}\left[h_{\mu\beta}h_{\nu}^{\beta}+\bar{g}_{\mu\nu}(ch_{\alpha\beta}^{2}+dh^{2})\right]\\
-\frac{1}{2}\bar{g}_{\mu\nu}\left[-\frac{5}{4}\bar{\nabla}^{\sigma}h^{\alpha\beta}\bar{\nabla}_{\sigma}h_{\alpha\beta}+\frac{1}{2}\bar{\nabla}_{\alpha}h_{\sigma\beta}\bar{\nabla}^{\beta}h^{\sigma\alpha}\right]\\
-\frac{1}{2}\bar{g}_{\mu\nu}\left[+\frac{1}{2}h^{\alpha\beta}\bar{\nabla}_{\alpha}\bar{\nabla}_{\beta}h-h^{\alpha\beta}\bar{\text{\textifsymbol[ifgeo]{48}}}h_{\alpha\beta}+\frac{1}{2}h\bar{\text{\textifsymbol[ifgeo]{48}}}h+\frac{3}{4}\bar{\nabla}^{\sigma}h\bar{\nabla}_{\sigma}h\right]\\
-\frac{1}{2}\bar{g}_{\mu\nu}\left[c(2-D)\bar{\text{\textifsymbol[ifgeo]{48}}}h_{\alpha\beta}^{2}+d(2-D)\bar{\text{\textifsymbol[ifgeo]{48}}}h^{2})-\bar{R}(ch_{\alpha\beta}^{2}+dh^{2})\right],
\end{multline*}
whose gauge-fixed version was given in the text.

\section*{Acknowledgments}
B.T. would  like to thank S. Deser for extended discussions on conserved charges in generic gravity theories and on the meaning of the vanishing charges. We would like to thank  D. Grumiller for a useful exchange on the log-modes of chiral gravity.

\end{document}